\documentclass[12pt,preprint]{aastex}

\shorttitle{The NLR of type 1 AGN}
\shortauthors{Xu et al.}

\newcommand{\kms}{km s$^{-1}$}
\newcommand{\oiii}{[\ion{O}{3}]}
\newcommand{\oii}{[\ion{O}{2}]}
\newcommand{\oi}{[\ion{O}{1}]}
\newcommand{\peroi}{\ion{O}{1}}
\newcommand{\caii}{\ion{Ca}{2}}

\newcommand{\sii}{[\ion{S}{2}]}
\newcommand{\nii}{[\ion{N}{2}]}

\newcommand{\fex}{[\ion{Fe}{10}]}
\newcommand{\fexi}{[\ion{Fe}{11}]}
\newcommand{\feii}{\ion{Fe}{2}}

\newcommand{\hii}{\ion{H}{2}}
\newcommand{\hb}{H$\beta$}
\newcommand{\hbb}{H$\beta_{\rm b}$}
\newcommand{\hbn}{H$\beta_{\rm n}$}
\newcommand{\han}{H$\alpha_{\rm n}$}
\newcommand{\ha}{H$\alpha$}
\newcommand{\LLedd}{${\rm L/L_{Edd}}$}
\newcommand{\LLeddbol}{${\rm L_{bol}/L_{Edd}}$}
\newcommand{\xLLedd}{${\rm L_X/L_{Edd}}$}
\newcommand{\n}{$n_{\rm e}$}
\newcommand{\te}{$T_{\rm e}$}
\newcommand{\cm}{cm$^{-3}$}

\slugcomment{ApJ accepted [00 00 2007]}

\begin{document}

\title{The narrow-line region of narrow-line and broad-line type 1 
Active Galactic Nuclei I. A zone of avoidance in density}

\author{Dawei Xu}
\affil{National Astronomical Observatories, Chinese Academy of
       Sciences, Beijing 100012, China; dwxu@bao.ac.cn}

\and
\author{Stefanie Komossa}
\affil{Max-Planck-Institut f\"ur extraterrestrische Physik,
Giessenbachstrasse 1, 85748 Garching, Germany; skomossa@mpe.mpg.de}

\and
\author{Hongyan Zhou} 
\affil{Department of Astronomy, University of Florida, Gainesville, FL32611, 
USA; and Max-Planck-Institut f\"ur extraterrestrische Physik,
Giessenbachstrasse 1, 85748 Garching, Germany;
and Center for Astrophysics, University of Science and 
Technology of China, Hefei, China; zhou@astro.ufl.edu}

\and
\author{Tinggui Wang}
\affil{Center for Astrophysics, University of Science and Technology of China,
       Hefei, China; twang@ustc.edu.cn}

\author{Jianyan Wei}
\affil{National Astronomical Observatories, Chinese Academy of
       Sciences, Beijing 100012, China; wjy@bao.ac.cn}

\begin{abstract}
The properties of narrow-line Seyfert\,1 (NLS1) galaxies,
the links and correlations between them, and the physics
behind them, are still not well understood. 
Apart from accretion rates and black hole masses,
density and outflows were speculated to be among the main drivers
of the NLS1 phenomenon.
Here, we utilize the diagnostic power of the \sii\ 
$\lambda\lambda 6716,6731$ intensity ratio to measure the density 
of the narrow-line region (NLR)
systematically and homogeneously for a large sample of NLS1 galaxies, 
and we perform a comparison with a sample of broad-line type\,1 
Active Galactic Nuclei (AGN). 
We report the discovery of a '{\em zone of avoidance}' in  
density in the sense that AGN       
with broad lines (FWHM(H$\beta$) $>$ 2000\,\kms) avoid low densities, 
while NLS1 galaxies show a wider distribution in the NLR density, 
including a significant number of objects with low densities.
A correlation analysis further shows that the Eddington ratio 
\LLedd\ anti-correlates with density. 
We investigate a number of different models  
for the '{\em zone of avoidance}' in density. 
Supersolar metallicities and temperature effects,
a strong starburst contribution in NLS1 galaxies, 
different NLR extents and selective obscuration are considered unlikely. 
Possible differences in the fraction of matter-bounded clouds
and differences in the interstellar media of the host galaxies
of NLS1 galaxies and broad-line Seyfert 1 (BLS1) galaxies 
can only be tested further with future observations.  
We tentatively favor the effects of winds/outflows, stronger
in NLS1 galaxies than in BLS1 galaxies, to explain the observations.  
\end{abstract}


\section{Introduction}
Optical and X-ray observations over the past few decades revealed a new
sub-class of AGN, termed Narrow-line Seyfert\,1 (NLS1) galaxies 
(e.g., Gaskell 1984; Osterbrock \& Pogge 1985). 
NLS1 galaxies are intriguing due to their
extreme emission line and continuum properties. Their optical broad
lines are narrower (FWHM \hb\ $\le$ 2000 \kms) than in `normal' 
broad-line Seyfert\,1 (BLS1) galaxies and they show strong \feii\
emission. Their X-ray spectra are sometimes, but not always, very soft
(e.g., Zhou et al. 2006, and references therein). Many emission-line
and continuum 
properties of AGN were found to correlate strongly with each other
(e.g., Boroson \& Green 1992, BG92 hereafter; Wang et al. 1996; 
Lawrence et al. 1997; Grupe et al. 1999; Vaughan et al. 2001; 
Xu et al. 2003; Sulentic et al. 2000, 2003; Grupe 2004).
The strongest variance, often referred to as `Eigenvector\,1' (EV1), is
defined by the correlation between the width of the \hb\ emission
line and the strength of the \oiii/\hb\ emission line ratio, and the
anti-correlation with the \feii/\hb\ ratio (e.g. BG92).
NLS1 galaxies are placed at one extreme end of EV1 
parameter space. The most common interpretation 
is that this regime is goverened by  
the highest Eddington accretion rates and/or lowest black hole
masses (e.g., BG92; Sulentic et al. 2000). 

Among other parameters%
\footnote{See \citet{stefanie06b} for a recent discussion. 
Briefly speaking,  other parameters considered to
be, perhaps, relevant in explaining NLS1 properties are
orientation (e.g., Osterbrock \& Pogge 1985; Puchnarewicz et al. 1992;
Collin et al. 2006; Zhang \& Wang 2006), the 
physics behind NLS1's radio properties (e.g. Komossa et al. 2006b),
metallicity (e.g., Mathur 2000; Komossa \& Mathur 2001; Nagao et al. 2002; 
Shemmer \& Netzer 2002; Romano et al. 2004; Fields et al. 2005), 
(ionized) absorption (e.g., Komossa \& Meerschweinchen 2000; 
Gierlinski \& Done 2004), 
and the galaxies' location on the $M_{\rm BH}-\sigma$ plane 
(e.g. Mathur et al. 2001; Wang \& Lu 2001; Grupe \& Mathur 2004;
Botte et al. 2005; Mathur \& Grupe 2005; Komossa \& Xu 2007)},
the density of an outflowing wind was 
firstly speculated to be a prominent driver of EV1 by 
Lawrence et al. (1997), given 
the connection of \feii\ strength with the presence of 
low-ionization, blueshifted broad absorption lines, 
and with blue-asymmetric emission lines. 
Winds and outflows play a crucial role in 
understanding the physics and evolution of AGN
(e.g., Elvis 2000, Hopkins et al. 2005),
and there is ample observational evidence for 
winds and outflows in AGN from
sub-kpc to galactic scale (see, e.g., Sulentic, Marziani \& Dultzin-Hacyan 2000;
Veilleux, Cecil, \& Bland-Hawthorn 2005, for recent reviews).
In enriching the nuclear environment with matter from the central region, 
winds may have an important impact on the 
gas densities in the emission-line regions. 

Regarding NLS1 galaxies, their high ratios of \LLedd\
 are likely particularly efficient in driving outflows. 
The large \oiii\ $\lambda$5007 blueshifts observed in some 
NLS1 galaxies are interpreted 
straightforwardly as the result of an outflow (e.g., Zamanov et al. 2002; 
Aoki et al. 2005, Boroson 2005). 
\citet{dewangan01} suggested that stronger outflows would likely push 
the broad-line region (BLR) further radially outward thereby resulting 
in narrower \hb\ lines
in NLS1 galaxies. They further speculated that the observed flux  ratios from 
the NLR and BLR can be explained in terms of density enhancements 
(see also Wills et al. 2000).
Outflows in NLS1 galaxies have been observed in 
terms of both blueshifted UV absorption lines 
(e.g., Laor et al. 1997a,b; Goodrich 2000) 
and UV emission lines (e.g., Leighly \& Moore 2004). 

Regarding the {\em density} of the emission-line regions
of NLS1 galaxies, only few previous estimates exist
(e.g., Wills et al. 2000; Kuraszkiewicz et al. 2000; Sulentic, Marziani \&
Dultzin-Hacyan 2000; Marziani et al. 2001; Bachev et al. 2004). 
Those which do exist, actually 
lead to partially conflicting results. 
While some sample studies \citep{kur00,wil00,mar01,bach04} suggested 
a high-density BLR in NLS1 galaxies, based on  
the large \ion{Si}{3}]~$\lambda1892$/\ion{C}{3}]~$\lambda1909$ ratios
measured from UV spectra
\footnote{One of the alternative explanations is 
that the carbon has been removed from gas 
phase by depletion onto dust grains \citep{cre02}.}, 
other inquiries favored low density emission-line clouds. 
\citet{rod00a} studied the emission-line properties of a sample of Seyfert\,1 
galaxies, including 7 NLS1 galaxies using optical and near-IR spectroscopy.
They found that the typical density of the \sii\ emitting zone is lower 
in NLS1 galaxies than in BLS1 galaxies.
\citet{rod97} tentatively favored a low-density BLR in NLS1 galaxies, inferred
from their photoionization modeling of the UV emission lines.
\citet{ferland89} derived densities similar to the canonical BLR value 
in strong \feii\ emitters, based on the ratio of the forbidden lines 
relative to the \caii\ triplet. 
\citet{baskin05} reported that the density of the \oiii\ $\lambda$5007 
emitting gas in the NLR decreases with steeper observed soft X-ray slope.

In addition, a few individual objects
were inspected more closely \citep{lao97b,lei04,veron04,veron06}.
In particular, analyses of the high S/N UV spectrum of the 
prototype NLS1 galaxy, I\,Zw\,1, indicated a high BLR 
density of 10$^{11}\,$cm$^{-3}$
and NLR density of $5\times 10^{5}\,$cm$^{-3}$ \citep{lao97b}. 
\citet{veron04} found on the same object 
from optical spectroscopy that the density 
of the low-ionization part of the NLR is of 
the order of 10$^{6-7}\,$cm$^{-3}$.

Given that few systematic measurements exist at all, 
and that those which do produce partially
conflicting results, it is important to explore this topic further.
We present for the first time a study of the NLR density 
for one of the largest homogeneously analyzed NLS1 samples to date 
and compare it with that of BLS1 galaxies. 
The electron density of the NLR can be measured by 
making use of the density-sensitive
line ratios
\oii\ $\lambda3729/\lambda3726$ and \sii\ $\lambda6716/\lambda6731$ (e.g., Osterbrock 1989).
In practice, the most important density diagnostic is the \sii\ line
ratio. Usually the density can not be directly inferred from the \oii\ line
ratio because the \oii\ $\lambda\lambda3726,3729$ doublet is often unresolved.
More indirectly, other line ratios  will also change with density.
For instance, a higher-density NLR would strongly boost
the \oi\ $\lambda$6300 line, thus
the intensity of \oi\ can be used to probe higher density regions 
(e.g., Komossa \& Schulz 1997; Barth et al. 2001).

This work is part of a series of papers investigating the properties of
the emission-line regions of type\,1 AGN, including NLS1 galaxies 
and BLS1 galaxies. 
In this first paper of the series,
we focus on topics related to density. Specifically, we attempt 
to answer the following key questions: 
(1) is there any difference in the NLR density between NLS1 galaxies 
and BLS1 galaxies?
(2) If so, do trends in density correlate with other parameters? and
(3) what are the key physical drivers to explain differences in the 
    NLR of NLS1 galaxies and BLS1 galaxies? 

This paper is organized as follows.
We present the data base and the sample selection in Sect.\,2.
In Sect.\,3, we describe our method of the optical spectral analysis.
The sample classification and an investigation
of selection effects is provided in Sect.\,4.
The key result, the detection of a '{\em zone of avoidance}' in the NLR density,
is reported in Sect.\,5. 
In Sect.\,6 we discuss the reality of the zone of avoidance,
followed by a discussion on its origin in Sect.\,7. 
We summarize our conclusions in Sect.\,8.

We use the terms NLS1 galaxies and BLS1 galaxies collectively for
high-luminosity and low-luminosity objects, i.e., NLS1 galaxies
for narrow-line type~1 quasars and for narrow-line type~1 Seyfert galaxies,
and BLS1 galaxies for broad-line type~1 quasars and for
broad-line type~1 Seyfert galaxies, respectively.
Throughout this paper, a cosmology with $H_{\rm 0}=70$\,\kms\,Mpc$^{-1}$,
$\Omega_{\rm M}=0.3$ and $\Omega_{\rm \Lambda}=0.7$ is adopted.

\section{Sample selection}
The combination of X-ray and optical observations has
proven to be an efficient way in detecting NLS1 galaxies over the last decades
(e.g., see Pogge 2000 and V\'{e}ron-Cetty,  V\'{e}ron \& Gon\c{c}alves 2001,
hereafter VVG01, for reviews).
The uniform optical spectroscopic galaxy survey known as the Sloan Digital Sky Survey 
(SDSS, York et al. 2000) is an excellent data base which enables us to study
AGN properties and, in particular, optical emission
line properties in a homogeneous way.

In this paper, in order to measure NLR densities,
we homogeneously analyze and compare optical emission line 
properties of a large number of NLS1 galaxies with BLS1 galaxies based on spectra obtained
in the course of the SDSS. We use the 3rd data release, DR3 \citep{sloandr3}.
The DR3 spectroscopic program covers an area of about 4188 deg$^2$. 
The two double spectrographs 
produce data covering a  wavelength range 3800--9200\AA\, 
at a spectral resolution $\approx 2000$. 
Exposure times for spectroscopy are determined such that a signal-to-noise (S/N)
of at least 4 pixel$^{-1}$ at $g=20.2$ is reached. 
The SDSS spectroscopic pipeline 
(Stoughton et al. 2002) 
performs spectral extraction, sky subtraction, removal of the atmospheric 
absorption bands, wavelength and flux calibration, and estimates the error 
spectrum. The processed DR3 spectra have not been corrected for Galactic 
extinction, but spectrophotometric calibration has been considerably improved 
since the First Data Release (DR1). We refer the reader to \citet{sloandr3}
for details of the changes.
  
\subsection{The NLS1 sample}
First, we extracted all NLS1 galaxies included in the the 11th 
edition of the "Catalogue of Quasars and AGN" (V\'{e}ron-Cetty \& V\'{e}ron 
2003; VV03 hereafter.). The defining criterion of NLS1 galaxies in VV03 is 
FWHM(\hb) $\le$ 2000 \kms. We further take into account only NLS1 galaxies 
with redshift $z$ less than 0.3. The redshift cut is imposed to ensure 
that \sii\ $\lambda\lambda$6716,6731 is observable, 
and furthermore, to ensure that the doublet is in a region 
free of noise from strong 
night-sky emission lines. The list of 309 selected NLS1 galaxies was then 
cross-correlated with the SDSS DR3, in order to obtain a homogeneous 
set of spectra of a large sample of NLS1 galaxies for 
uniform spectral analyses. 
This procedure resulted in the selection of 119 sources. 
In order to get an accurate density measurement, we 
require \sii\ $\lambda\lambda$6716,6731 to have S/N 
greater than 5. We also removed a handful of objects with spectral defects 
(e.g. problems with sky subtraction, or missing signal over a range of 
wavelengths; see Strateva et al. 2003) by visual inspection. 
This leaves a total of 58 objects.

\subsection{The BLS1 sample}
Recently, Boroson (2003) presented a sample of 107 low-redshift 
(redshift $z<0.5$) type~1 AGN from the SDSS Early Data Release (EDR, Stoughton 
et al. 2002)%
\footnote{We do not divide the Seyfert 1 class into subclasses such as 
Seyfert 1.5-1.9 (as defined by \citet{ost85}), since the definitions of 
Seyfert classifications depend on the resolution of the spectra used and 
the noise in the spectra \citep{gru99,good89}.}.  
82 out of these objects match our redshift constraint, i.e., $z<0.3$.
We use this sample to build the control BLS1 sample imposing the same 
redshift cut and S/N limit of \sii\ $\lambda\lambda$6716,6731 (i.e., $z<0.3$ 
and S/N $>5$) as to the NLS1 sample.   
The processed DR3 spectra were downloaded for spectral inspection in 
accordance with the data set of the NLS1 sample.
With these restrictions, 48 objects survive, 
13 of which overlap with the NLS1 sample. 

\section{Spectral Analysis}
The optical emission-line properties (line widths and line ratios) are among
the defining criteria of the NLS1 phenomenon. The exact and homogeneous 
measurements of the emission-line parameters of AGN provide us with basic
knowledge about emission-line regions, and enable us to 
investigate the relationship between NLS1 galaxies and BLS1 galaxies. 
However, the
published data were compiled in a heterogeneous way. The use of different line
profiles (e.g., Lorentzian vs Gaussian profiles) leads to strong differences in
the \hb\ widths (e.g., VVG01).
In addition to the Lorentzian and Gaussian representations of line profiles,
a direct measurement of the width at half of their maximum intensity
is sometimes adopted 
(e.g., Williams, Pogge \& Mathur 2002; Boroson et al. 2003).
Also, the strong \feii\ contamination makes it difficult to
measure \hb\ and brings large uncertainties in determining the \hb\
line width. 

Here we perform a homogeneous spectral analysis for the NLS1 and BLS1 samples. 
The objects are then re-grouped into the NLS1 or BLS1 sample for further 
investigation, according to the widths of the broad \hb\ line (Sect.\,3.2) 
obtained in our spectral analysis. 

In a first step, the SDSS spectra were corrected for 
the Galactic extinction using the reddening map 
of Schlegel et al. (1998), 
and then shifted to their rest wavelength, adopting the SDSS value 
of the redshift from the header of each spectrum.
Once these steps were completed, we removed the stellar continuum, 
subtracted the \feii\ complexes and performed the spectral analysis for the 
emission lines following the procedures described below.

\subsection{Subtraction of starlight and nuclear continuum}
SDSS spectra are acquired with a pair of fiber-fed spectrographs. 
Each fiber subtends a diameter of 3$''$, corresponding
to $\sim 6.5$~kpc at $z=0.1$. This aperture is large enough
to let through not only the emission from the nucleus, but also
a substantial amount of starlight from the host galaxy (e.g., Vanden
Berk et al. 2001, Hao et al. 2005). The accurate removal of the stellar
contribution is essential to reliably measure the emission-line
spectrum, such as the line widths and line strengths. Particularly,
the reliable classification of NLS1 galaxies is strongly dependent on the
measurement accuracy of the width of the broad \hb\ line. Furthermore, in
many of the spectra there is a clear contribution from blends of
\feii\ line emission. Well-studied NLS1 galaxies usually show strong optical
\feii\ emission features on both the blue and red sides of the
\hb\--\oiii\ complex (e.g., V\'{e}ron-Cetty \& V\'{e}ron 2003). 
In order to reliably measure line
parameters, we choose those wavelength ranges as pseudo-continuum,
which are not affected by prominent emission lines, and 
then decompose the spectra into the following 4 components (see
Zhou et al. 2006 for details):
\begin{itemize}
    \item A starlight component modeled by 6 synthesized galaxy
    templates, which were built from the synthetic spectral library of Bruzual
    \& Charlot (2003) using the algorithm of Ensemble Learning for
    Independent Component Analysis (EL-ICA, Lu et al. 2006).
    These templates were broadened by convolution with a Gaussian to match the
    stellar velocity dispersion of the host galaxy.
   \item A power-law continuum to describe the emission from the active nucleus.
    \item An \feii\ template obtained by V\'{e}ron-Cetty \& V\'{e}ron (2003).
    This template covers the wavelengths between 3535$-$7534 \AA,
    extending further to both the blue and red wavelength ranges than
    the \feii\ template used in BG92. This makes it more advantageous
    in modeling the \feii\ emission in the SDSS spectra. We assume
    that \feii\ has the same profile as the broad component of \hb\
    (see the next subsection).
   \item A Balmer continuum generated in the same way as Dietrich et al. (2003).
\end{itemize}

The modeling is performed by minimizing the reduced $\chi^2$ in the
fitting process. The final multi-component fit is then subtracted
from the observed spectrum.
An example of the residual spectrum is plotted in Fig.\,1.

\subsection{Decomposition techniques}
The multicomponent-subtracted spectra are used to measure the 
non-\feii\ line properties. 
The broad Balmer lines in AGN exhibit a wide variety of profile shapes and a 
large range in width \citep{ost82,de85,cre86,str91,mil92,veron01}, and they 
are often strongly asymmetric \citep{cor95}. In many cases the Balmer lines
are mixtures of broad and narrow components. Differences in the relative 
strengths of these components account for much of the diversity of broad 
line profiles \citep{fra92,wil93,bro94,cor95,cor97,veron01}.
A proper decomposition of the NLR and BLR line emission contribution of 
the Balmer lines is of great importance to address the 
physical properties of the emission-line regions. 
Particularly, the width of the broad \hb\ profile has significant 
impact on the reliable classification of a galaxy as a NLS1 galaxy. 

In order to measure the parameters of the BLR emission lines, 
the NLR line emission
contribution has to be removed first. Using homogeneous sets of spectra, 
previous studies (e.g., Filippenko \& Sargent 1988; Ho et al. 1997; 
Greene \& Ho 2004; Zhou et al. 2006) have shown that the narrow Balmer emission
profiles are well matched to those of \sii\ $\lambda\lambda$6716, 6731. 
Moreover, the widths of \sii\ and \nii\ doublets trace the stellar velocity 
dispersion of galaxy bulges better than that of \oiii\ within the 
uncertainties (Greene \& Ho 2004). Therefore, we use
the strong profiles of \sii\ $\lambda\lambda$6716, 6731 (those with S/N $>$5) 
as a NLR template profile for the narrow component of the Balmer lines% 
\footnote{Occasionally, we also make use
of spectra with weak or absent \sii\ emission. These are
used to measure \oi\ (see Sect. 5.3). 
In such a case, the \oiii\ profile is used as a substitute for \sii.}.

To characterize the NLR emission-line profiles, we fit these lines using
Gaussian profiles. Most \sii\ $\lambda\lambda$6716,6731 and 
\nii\ $\lambda\lambda$6548,6583 lines can be well fit employing one single
Gaussian profile% 
\footnote{ Using spatially-resolved {\em HST} spectra, Rice et al. (2006) 
identified both blue and red asymmetries in some [S II] line profiles, 
which are primarily due to nuclear line-emitting gas, rather than more 
symmetric emission from the NLR on larger scales.}.
A large fraction of the \oiii\ $\lambda\lambda$4959,5007 lines show strongly 
asymmetric profiles. In those cases, a second component is then added 
to represent the line wings. When we fit \sii\ $\lambda\lambda$6716,6731,
\nii\ $\lambda\lambda$6548,6583 and \oiii\ $\lambda\lambda$4959,5007 lines, 
the separation of the lines of each doublet is fixed to the laboratory value. 
Each pair is assumed to have the same profile%
\footnote{The line width of each component of the doublet may be slightly 
different due
to stratification in the NLR. However, Rice et al. (2006) found the widths 
of the two lines differ by less than 3\%, which is within the error of 
the Gaussian fit parameters and hard to discern with data of moderate S/N.}. 
While the flux ratios of the \nii\ and \oiii\ doublets 
are fixed to the theoretical ratio of $3:1$, the intensity ratio of the \sii\ 
doublet is measured, and then used to derive the density.

We used the IRAF package SPECFIT \citep{kriss94}
to measure blended lines and separate the NLR from the BLR emission. 
The actual profile shape of the broad Balmer lines of 
NLS1 galaxies is still an issue debated. While the broad Balmer line component 
can be well represented by 
Gaussian profiles, particularly by a 
combination of multiple Gaussian components
\citep{rod00b,nagao02,xu03,greene05a,greene05b,dietrich05}, successful 
fitting can also be accomplished with a single Lorentzian profile 
\citep{lei99,veron01} for some NLS1 galaxies.  
\citet{veron01} suggested that the broad Balmer lines of 
NLS1 galaxies were better fitted with a single Lorentzian profile than 
a single Gaussian profile (see also Sulentic et al. 2002), since many NLS1 
galaxies were located in the H\,II region in their diagnostic diagrams if a 
single Gaussian profile was adopted. However, as noted in \citet{eva88}, 
the choice of Gaussian or Lorentzian profiles as representatives of the 
observed emission lines may bear no physical meaning. Furthermore, the 
broad-line profiles show complex shapes and asymmetries that can not be 
described with a single component, indicating the presence of two 
or multiple components, 
independent from the type of profile used to fit \citep{dietrich05}. 

In order to isolate the narrow and broad components of the Balmer lines, 
we fix the width of the narrow component (determined from the width of \sii\
as described above) and only leave its strength as fit parameter.  
The broad part of the line profiles we fit by using a combination of 
two Gaussian profiles, as well as a single Lorentzian profile. 
For the approach with a multi-Gaussian components fit, we measure the FWHM of 
the profile and its integrated flux from the final combined model for 
the broad component. The individual Gaussian components have no physical 
significance by themselves but are only used to serve as a description of 
the complex line shapes of the broad components as far as the data 
quality allows. 
While comparable and equally reasonable results can be achieved with both
single Lorentzian and two Gaussian profiles for most broad components
of the Balmer lines of NLS1 galaxies, 
generally no acceptable fit is possible when 
employing the Lorentzian profiles to fit BLS1 galaxies. 
We will use the results of the multiple Gaussian fit for NLS1 galaxies 
for classification and further correlation analysis, for its simplicity, 
and furthermore, for a direct comparison with the BLS1 control sample and  
with previous studies (e.g. BG92; Grupe et al. 1999; Vaughan et al. 2001).
However, for comparison purposes, we still report results from
Lorentzian fitting in our key figures (Fig. 5, 6). 
We further checked the location of NLS1 galaxies and BLS1 galaxies in the 
diagnostic diagrams using the standard  
emission-line ratios. 
In contrast to Fig.\,8 of VVG01, 
most of our NLS1 galaxies locate in the AGN regime
for both Gaussian and Lorentzian fits. 

The broad component of \hb\ is referred as \hbb, and the narrow 
component as ${\rm H\beta_n}$. 
The dominant uncertainties in the line parameter measurements 
can result from the continuum subtraction and component decomposition. 
The latter depends on the line profile shape.
Although the line measurements of Gaussian and 
Lorentzian fits are precise, they
are only true estimates if the lines can be correctly represented by the 
profiles. The average error in the width of narrow lines is of approximately 
5\%. Errors in the width of broad lines, introduced by the
different profile types, e.g., two Gaussians or a single Lorentzian, 
are about 10\%. For most objects, the uncertainties of flux measurements 
of emission lines are less than 10\%. 
The typical measurement error of the \sii\ doublet is about 5\% for our sample.

\section{Sample re-classification and selection effects}

\subsection{Re-classification of the sample} 
In order to follow the standard practice to distinguish between 
NLS1 galaxies and BLS1 galaxies by FWHM(\hbb), we re-classified the
objects of our sample accordingly, after having carefully measured
FWHM(\hbb).
A total of 55 objects with
FWHM(\hbb) $\le 2000$ \kms are included in the final NLS1 sample, 
while 39 with FWHM(\hbb) $>$ 2000 \kms\ are included in the BLS1 sample.

\subsection{Luminosity and redshift distributions}
As described in Sect.\,2, we do not set any selection criteria upon our 
samples, which combines data from various different sources and thus
is not statistically complete. It is therefore necessary to examine 
whether or not the NLS1 and BLS1 samples are drawn from comparative 
populations such that 
luminosity effects (in particular, the number of quasars vs Seyferts in each
sample) and evolutionary effects can be excluded.
In order to check the possible biases between the two samples, we look into
the absolute magnitude distribution and the redshift distribution.

First, we investigate whether or not there is a systematic difference in the 
luminosity distribution between the two samples. Here we trace the luminosity 
with the absolute $i$\,magnitude, which is calculated from the $i$\_psf 
magnitude (e.g. Schneider et al. 2005; Vanden Berk et al. 2004), 
by correcting the $i$\,measurements for Galactic extinction 
(Schlegel et al. 1998) and assuming 
a powerlaw spectral energy distribution (SED)
($f_v \propto \nu^\alpha)$, where $\alpha=-0.5$ (e.g., 
Vanden Berk et al. 2001). 
Following the previous studies of the SDSS AGN (Schneider et al. 2003, 2005; 
Nagao et al. 2005), we define objects that have luminosities larger than
M$_i=-22.0$ as quasars. 
We show the histograms of the absolute $i$\,magnitudes 
for the NLS1 galaxies and BLS1 galaxies
in Fig.\,3. We apply the Kolmogorov-Smirnov (K-S) statistical test to check 
how closely the distributions resemble each other.  
The resultant K-S test probability is 0.33, which means 
the two distributions are statistically indistinguishable.

Secondly, we explore the redshift distribution. The histograms of the 
sources' redshifts are shown in Fig.\,3. The NLS1 and BLS1 samples 
possess similar average 
redshifts. The value and 1$\sigma$ deviations for 
the two samples are $0.128 \pm 0.068$ and $0.142 \pm 0.066$, respectively.
Moreover, since both samples are compiled from low-redshift objects with 
$z<0.3$, no strong redshift bias is expected, anyway.
We conclude that our two samples do not show significantly different
luminosity and redshift distributions.

\subsection{Optical \feii\ emission versus \oiii\ emission}
In this section we wish to demonstrate that our (NL)S1 sample follows
the same correlations found previously for (NL)S1s as a class, and 
therefore, that our NLS1 galaxies are representative for the NLS1 population as 
a whole, as far as previously known correlations are concerned.
Among optical emission lines of NLS1 galaxies and BLS1 galaxies, 
the most striking correlation is the so called EV1 
(e.g., BG92; Grupe et al. 1999; Sulentic et al. 2002, 2003; Xu et al. 2003), 
i.e., that \oiii/\hb\ emission 
is weak in objects with strong \feii\ emission and vice versa. NLS1 galaxies 
are at the extreme end of this correlation, often 
showing strong \feii\ emission and weak \oiii/\hb$_{\rm total}$ emission. 

The optical \feii\ strength, R4570, is defined as the ratio of the flux of 
the \feii\ complex between the rest wavelength 4434\AA\, and 4684\AA\, to 
the total \hb\ flux, including the narrow component (e.g. BG92, VVG01).  
The average values of R4570 for our NLS1 and BLS1 samples are 0.75 and 0.30, 
respectively. The regime of high values of R4570 is solely 
occupied by NLS1 galaxies. 
The results are consistent with previous studies (e.g., Rice 
et al. 2006 and references therein). 
In Fig.\,4, we plot R5007, the ratio of \oiii$\lambda5007$ flux to the total 
\hb\ flux, as a function of R4570. Both emission lines span almost the 
whole range from the low to high end observed in 
NLS1 galaxies and BLS1 galaxies (e.g., 
Fig.\,11 in VVG01), suggesting there is no preferential selection effect 
to objects with extreme properties. Moreover, an anti-correlation between 
the two parameters is present, with a correlation coefficient 
r$_{\rm s}=-0.37$. The probability of the null correlation is
P$_{\rm null}<10^{-3}$. 
These quantities are calculated from the Spearman rank-order 
correlation analysis. Hence, 
the NLS1 and BLS1 samples under study can be considered as representative
samples of these two subclasses, though they are not statistically complete.

\section{A zone of avoidance in the NLR density}
\subsection{Measurement of density}
We use the {\em density} diagnostic \sii\ $\lambda6716$/$\lambda6731$ 
to measure the NLR electron density.
The electron density is calculated following \citet{ost89} using 
the task {\em temden} of the IRAF package STSDAS \citep{shaw94}. 
This approach comes with two limitations.
Firstly, around densities of 10$^3$--10$^4$\,\cm\ we approach the critical densities
of \sii\,$\lambda6716$~($1.5 \times 10^3$\,\cm) 
and \sii\,$\lambda6731$~($3.9 \times 10^3$\,\cm).
Secondly, the \sii\ line ratio also depends on temperature. However,
the dependence is only weak in the range of temperatures of the regions studied.
We fix the electron temperature at \te\,$=10,000$K, typical for photoionized 
gas in the NLR, and comment on possible temperature effects later.
The \sii-based density measurement allows us to probe the \sii-emitting
part of the NLR. In order to get a first impression also on the 
high-density part of the NLR, later we make use of the strength of
the \oi\ line (Sect. 5.3){\footnote{We note that the NLR is most likely
composed of  
emission-line clouds with a range in densities 
(e.g., Komossa \& Schulz 1997; Peterson 1997; 
Rodriguez-Ardila, Pastoriza \& Donzelli 2000a; Brinkman et al. 2000),
and that NLR density increases  
toward small radii (e.g., 
Fraquelli, Storchi-Bergmann \& Binette 2000; Bennert et al. 2006a,b).}}. 

The [SII] ratio is only a good density diagnostic in a certain range of 
densities because depopulation of the upper levels changes from being primarily 
radiative at low densities to primarily collisional at high densities. 
The {\rm low-density} limit is $\sim 10$\,\cm,
while for densities higher than $10^{4}$\,\cm, collisional de-excitation
becomes more and more important. Most of our sources distribute in a regime 
where the methods are well applicable.
In the {\rm low-density} limit (\n\,$< 10$\,\cm), the error
in density is larger due to the saturation of the relation 
between the line ratio and electron density, so these derived 
densities should be treated with caution. 
An \sii\ ratio of 1.42 corresponds to a density of $\sim 10$\,\cm.
Only two objects, SDSS J011448.68-002946.1 and SDSS J161809.38+361957.8, 
have ratios greater than 1.42, which places the sources at the 
low end in the density distribution.

\paragraph{Notes on individual objects.}
Among the whole sample, only objects which are above a certain S/N ratio
in the \sii\ line 
were kept for further analysis (Sect.\,2). Among these, we then individually
re-inspect all objects below a density threshold of 
\n\,$=140$\,\cm (log\,\n\,$=2.15$\,\cm, 
i.e., those in regime "A" in Fig.\,5) 
to check for peculiarities in 
spectral features, and to check the robustness of spectral fitting and 
thus reliability of the density estimate in the 
{\em low-density} regime.  

We comment here on those objects 
which are extreme in the \sii\ ratio, FWHM or other spectral features: 
(1) There is one single broad-line object actually located in the 
'{\em zone of avoidance}' in density. This is SDSS J011448.68-002946.1. While
its \sii\ ratio is extreme (R(\sii)$=1.47 \pm 0.03$), and within the 
errors beyond
the '{\em low-density}' regime, its other emission lines are 
not unusual. The profile of \sii\ $\lambda6731$ deviates somewhat from
a Gaussian, but there is nothing else very peculiar about it. 
This is the only outlier in Fig.\,5. 
(2) The low-density NLS1 SDSS J092247.03+512038.0 is peculiar in that 
the peak of the \oiii\ $\lambda5007$ line is blueshifted with respect to the 
low-ionization forbidden lines (e.g., \oii, \nii\ and \sii) and the narrow
component of \hb\ by more than 400~\kms. The large velocity shift places 
it among the \oiii\ blue outliers \citep{zamanov02}.
Moreover, \oiii\ $\lambda5007$ is almost as broad as \hbb\ 
(FWHM 1060 \kms\ and 1250 \kms\ for \oiii\ $\lambda5007$ and \hbb,
respectively.), while the low-ionization forbidden lines such as \sii\
are narrow (FWHM(\sii) $=220$~\kms). It is a strong \feii\ emitter
with ${\rm R4570}=1.32$. 
(3) SDSS J161809.38+361957.8 is the object with the most extreme
combination of \hbb\ line width 
(FWHM(\hbb) $=700$~\kms\ for a Lorentzian profile fit) and density 
(\n\,$=2.0$\,\cm). 
The emission lines have a very wide range of excitation, the highest
corresponding to coronal lines of \fex\ $\lambda6374$ and \fexi\ $\lambda7892$.
We also detect the broad low-ionization line \peroi\,$\lambda8446$, which 
is generally produced in a region with very high density 
(e.g. Komossa \& Bade 1999).  
It is the only low-density source of our sample which has 
the \peroi\,$\lambda8446$ line detected.         
(4) The low-density NLS1 SDSS J120226.76-012915.3 shows
exceptionally strong \feii\ emission of ${\rm R4570}=2.74$.
It is the most extreme \feii\ emitter in our sample.
(5) Finally, we mention that among the high-density sources,
the spectrum of 
the BLS1 SDSS J013527.85-004448.0 
is special in that it shows the highest ratio of \oii/\oiii\ of 
the whole sample (\oii/\oiii$_{\rm obs} = 1.6$) 
{\footnote{More detailed account on multi-wavelength properties
and on (unusual) line profiles and line ratios of individual sources 
will be given in a follow-up paper (Xu et al. 2007, in prep.)}}.  

\subsection{Density versus FWHM(\hbb): a zone of avoidance}
One of our main goals is to examine whether or not there is a difference in
electron density \n\ 
between NLS1 galaxies and BLS1 galaxies, in order to test different NLS1 models.
The \sii\ $\lambda6716/\lambda6731$ line ratio in our sample ranges
from 0.87 to 1.47.
We plot the ratio versus the FWHM of \hbb\ in Fig.\,5.
The typical error on the ratio is $\sim$0.06.
The histograms of the ratios for the NLS1 and BLS1 samples are also shown.
NLS1 galaxies show 
\sii\ ratios in the range from 0.94 to 1.43. 
17 out of 55 NLS1 galaxies have a ratio higher than 1.28,
while only one out of 39 BLS1 galaxies has a high ratio (i.e., 1.47). 
The other 38 BLS1 galaxies occupy the range from 0.87 to 1.27.
The average ratios and 1$\sigma$ deviations of the two samples are
$1.23 \pm 0.12$ and $1.12 \pm 0.10$, respectively.

In Fig.\,5 (lower panel), we display the inferred electron density 
against the FWHM of \hbb. We find that the sources do not homogeneously 
populate the \n\,--FWHM(\hbb) diagram.
The key detection is a '{\em zone of avoidance}' in the diagram.
While the 38 BLS1 galaxies avoid low average densities, and all show 
\n\,$>140$\,\cm\ (regime\,C), 
NLS1 galaxies show a larger scatter in density 
in the range \n\,$=2 \sim 770$\,\cm, 
including a significant number of objects with low densities. 
17 out of the 55 NLS1 galaxies under study show 
\n\,$<140$\,\cm\ (regime\,A) and are 
clearly separated from the range occupied by BLS1 galaxies. 
The other 38 NLS1 galaxies overlap well with the range in density 
for BLS1 galaxies (regime\,B)% 
\footnote{Results are robust, no matter whether a Gaussian or Lorentzian 
profile is used for the broad component of \hb. 
A larger scatter, rather than a strict cut-off in density 
for BLS1 galaxies might  
appear if the sample size increases significantly, 
but we expect our findings do still hold on average.}. 
The average electron densities for zone A, B and C are 69, 294 and 380\,\cm, 
respectively. 
We apply the K-S statistical test 
on the distributions of density for NLS1 galaxies vs BLS1 galaxies.
The resultant K-S test probability is 0.0002, which means
that the two density distributions are significantly different.

\subsection{\oi\ emission versus FWHM(\hbb)}
In order to probe also the high-density NLR regime, well above the critical 
densities of the two \sii\ lines, we concentrate on the emission line \oi\ $\lambda6300$. 
Even though the ratio \oi/\ha\ is also influenced by other parameters, 
\oi\ is strongly boosted for high densities (e.g., Fig.\,4 and 8 of 
Komossa \& Schulz 1997) and we thus checked for any trends
and correlations between \oi\ intensity and FWHM(\hbb)
(Fig.\,6), which we regard as a supplement to  the \n\,--\,FWHM(\hbb) 
diagram (Fig.\,5). 
In Fig.\,6, we plot the ratio \oi\ $\lambda6300$/\han\
 as a function of the FWHM(\hbb).
An anti-correlation between the two parameters is found with a
correlation coefficient r$_{\rm s}=-0.34$, and a probability of the null
correlation P$_{\rm null}<10^{-2}$, calculated from the Spearman
rank-order correlation analysis. The trend meets our expectation 
from Fig.\,5, in the sense that
BLS1 galaxies show, on average, higher \oi/H$\alpha$ than NLS1 galaxies. 

\section{Reality of the zone of avoidance}
The sources of our sample do not homogeneously populate the \n\,--\,FWHM(\hbb) 
diagram, but show a '{\em zone of avoidance}', in the sense that 
BLS1 galaxies lack low average NLR densities. 
Is there any data analysis or selection effect that could cause a spurious 
{\em zone of avoidance} in the BLS1 galaxies regime, or could mimic a larger 
density scatter in the NLS1 galaxies regime? We discuss and reject several 
possibilities in turn.

\paragraph{Magnitude distribution.}  
Firstly, we note that the magnitude distributions of the NLS1 and the BLS1 
sample are similar (also in dependence of redshift; Fig.\,3).
We note that both extremes of the distribution in the \n\,--\,FWHM(\hbb)
diagram include both quasars and Seyferts (but more Seyferts, 
since the total number of Seyferts is higher than the quasars).

\paragraph{Profile shape.}
Secondly, our results are robust, independent of the profile the
broad emission lines are fit with, either a Lorentzian profile or a 
combination of two Gaussian profiles (see Fig.\,5). 

\paragraph{Atmospheric absorption effects.} 
We then checked whether the density estimate for the small-FWHM regime 
is more unreliable (thus, a larger scatter) because at a certain redshift 
range ($z=$\,0.130--0.140), the atmospheric O$_2$ absorption line at 7620\AA\
overlaps with one or the other sulfur line, making line estimates 
more unreliable (even though atmospheric absorption is generally already 
corrected for the released SDSS spectra).%
\footnote{The other atmospheric O$_2$ absorption line at 6870\AA\ will not 
overlap the sulfur lines, given the redshift range 0.034--0.289 of the
objects in our sample.} 
In that case, sources in the regime with the larger range in densities 
(zones A \& B in Fig.\,5) should have a specific, narrow redshift range. 
This is not the case, however. The low density objects vary in redshift 
between $z=$\,0.034--0.280, while the rest of the sample 
shows $z=$\,0.038--0.289.  

\paragraph{Signal/Noise.}
In order to get an accurate \sii\ line ratio and thus density measurement,
we required \sii\ $\lambda\lambda$6716,6731 to have S/N greater than 5
which is 
quite a common cut-off imposed on SDSS data.
90\% of our spectra are actually of S/N (\sii) $\ge$ 10. 
In order to see whether the S/N of our spectra affects in
any kind of way our measurement of the [SII] ratio, we
have checked whether one class of objects (those with low density) or the other class
(with high density) shows an excess of sources 
with low S/N. 
However, this is not the case. 
We have run a Spearman rank-order correlation analysis on 
the distribution of the \sii\ ratio in dependence of S/N, 
and we do not find any correlation (r$_{\rm s}=-0.05$, P$_{\rm null}=0.63$).

\paragraph{Faint broad wings in the Balmer lines.}
In case of very broad Balmer lines, if the broad wings of the line are over- or
underestimated, then a corresponding small under- or over-estimate in the \sii\ lines
might result. The \sii\ $\lambda$6716 line might then be slightly more strongly 
affected, and that might marginally change the measured \sii\ ratio.  However, in 
most of our sources the [SII] lines are well separated from faint broad wings (see Fig.\,2)

\paragraph{High and low density limits.}
\sii\ $\lambda$6716/$\lambda$6731 is only a good density 
diagnostic in a certain 
range of densities. The low-density limit is ${\rm \sim 10}$\,\cm,
while for densities higher than $10^{4}$\,\cm, collisional de-excitation 
becomes more and more important. Most of our sources distribute in a regime 
where the methods are applicable. Since the conversion of the \sii\ ratio 
into density is non-linear, the estimate of density in the low-density regime 
(Fig.\,5.3 of Osterbrock 1985) comes with larger errors. These are shown in 
Fig.\,5. Within these errors, objects with small FWHM(\hbb) still have, 
on average, lower densities. 

The whole diagram (i.e., across the whole FWHM(\hbb) range), may, to a minor extent, 
be biased toward excluding very high densities by the original selection effect 
of using only spectra with significant S/N in the \sii\ lines. This 
pre-selection cannot be avoided. 
However, since \sii\ is collisionally de-excited for densities well above 
$10^4$\,\cm, the pre-selection criterion then is prone to excluding 
objects with extremely high densities, should they exist at all.
    
\paragraph{Influence of temperature.}
The estimate of density using the ratio \sii\ $\lambda$6716/$\lambda$6731 also 
depends on temperature. As described in Sect.\,5.1, we fix \te\,$=10000$\,K. 
An estimate of the temperature is possible with the temperature-sensitive
emission-line ratio \oiii\ $\lambda$4363/\oiii\ $\lambda$5007.
However, \oiii\ $\lambda$4363 is a faint line which generally
comes with large measurement errors. 
We have inspected the highest-quality spectra available in the 
\oiii\ $\lambda$4363 wavelength range (high S/N and/or easy deblending of 
H$\gamma$ and \oiii); both, among the lowest density object, and, for 
comparison, a similar number among the higher density objects. We do not 
find systematically higher temperatures for the low-density objects. 
We further note that given the scaling of density with temperature, 
$n_{\rm e}(T)=n_{\rm e}({\rm obs}) \times  \sqrt{T/10\,000}$ \citep{ost89}, 
even an increase
in temperature up to 40,000K (close to the average of four Sy1 galaxies
of Bennert et al. 2006b), would only change density by a factor of 2, much 
less than the scatter in the observed density values. We come back to this 
point in Sect.\,7.2. 

\paragraph{Starburst contribution.}
If there was a strong systematic starburst/\hii\ contribution to the 
emission lines, 
then we would expect the density estimates to be biased 
toward lower values, more 
typical for \hii\ regions.   
Since our sample shows unambiguous AGN-like line ratios, 
we expect the vast majority of the emission to come from the NLR, 
rather than star-forming regions. 
We further use the star-forming indicator, \oii\ $\lambda$3727 
(e.g, Hippelein et al. 2003; Kewley \& Geller 2004; Ho 2005), 
to track the starburst contribution to the optical   
emission lines of the AGN of our sample. 
\oii\ $\lambda$3727 is prominent in \hii\ regions,  
while the ratio \oii/\oiii\ is observed and predicted to be relatively 
weaker in Seyfert galaxies.
We compare the line ratios \oii\ $\lambda$3727/\oiii\ $\lambda$5007 
of NLS1 galaxies and BLS1 galaxies, in order to check whether or not the 
\oii\ emission is on average stronger in NLS1 galaxies.
In Fig.\,7 we show the distributions of the ratio 
prior to reddening correction and after reddening correction%
\footnote{For dust reddening correction we use the average reddening curve
of Osterbrock (1989, his Table\, 7.2) and an intrinsic value of 
\ha$_{\rm n}$/\hbn\,=\,3.1.}. 
No difference is present between the \oii/\oiii\ ratio for NLS1 galaxies and BLS1 galaxies. 
Thus, we conclude that the starburst contribution
is not a possible explanation for the \n\,--\,FWHM(\hbb) distribution.
The average observed \oii/\oiii\ ratios for the NLS1 and BLS1 sample are
0.32 and 0.29, while the average reddening corrected ratios are 0.35 and 0.33,
respectively.

\paragraph{Spatially resolved NLRs and viewing angle effects.}
Finally, we checked whether nearby objects might be spatially resolved, 
such that the low-density outer part of the NLR is missed in the fiber
which would then result in an overestimate of the density in
these objects.  
We find that the fiber diameter of 3$''$ corresponds to 
$\sim 2$\,kpc at ${\rm z=0.034}$ (the lowest redshift in our sample), 
while the gas in the NLR is typically distributed over a distance 
$r \sim 10-1000$\,pc from the nucleus. 
Therefore, SDSS fibers would always cover the entire NLR.

\section{On the origin of the zone of avoidance}

The key result of this study is the detection of a {\em zone of avoidance} in 
the density - FWHM(\hbb) diagram (Fig.\,5): BLS1 galaxies (FWHM(\hbb) $>$ 2000~\kms) 
avoid low average densities, and all show \n\ $>140$\,\cm. 
On the other hand, NLS1 galaxies show a larger scatter in densities in the range 
\n\,$=2 \sim 770$\,\cm, including a significant number of objects with 
low densities. The results obtained for \sii\ are consistent with 
the \oi\,--\,FWHM(\hbb) diagram as shown in Fig.\,6, 
which shows lower average \oi/H$\alpha$ in NLS1 galaxies.
In the following, we first confront our results on density with predictions
or indications on density effects of existing NLS1 models, and then discuss
further possibilities.

\subsection{Coupling between NLR and BLR?}

Given occasional reports of signs of lower or higher-than-average
BLR density of NLS1 galaxies (e.g., Kuraszkiewicz et al. 2000; 
Rodriguez-Ardila et al.2000a; Marziani et al. 2001;
Komossa \& Mathur 2001; Xu et al. 2003; Bachev et al. 2004)
we then also expect the same trend to reflect in the NLR densities
-- {\em if} the properties of NLR and BLR are closely   
linked{\footnote{The exact relation between BLR and NLR in AGN is still
under examination. While some models do predict a close link, others
assume or indicate that both cloud components are of different and
unrelated origin. Models that argue against a common link include indications
that NLRs are just normal interstellar medium (ISM) in the host galaxy, 
while the BLR has separate
origin. Also, the fact that the FWHMs of the Balmer lines of BLR and NLR do not
generally correlate (e.g., BG92; Grupe et al. 1999; Vaughan et al. 2001; 
Xu et al. 2003) 
suggests different kinematic components. 
On the other hand, there are models and observations that do predict
a link, including common wind outflows (e.g., Schiano et al. 1986); the
suggestion that the apparent gap between BLR and NLR is solely caused by
dust effects (Netzer and Laor 1993);
and observational links between BLR and NLR parameters, in particular,
between line strength, line asymmetry and the shift of the
line centroids 
(e.g., Xu et al. 2003).}}. 
Such models can then be tested by measuring the NLR density.
We start with a short review of previous measurements,
then come back to our new results. 

Few NLS1 galaxies have been studied with respect to the density of
their NLR, so far.
\citet{rod00a} reported NLR density measurements of seven NLS1 galaxies,
based on the \sii~$\lambda6716$/\sii~$\lambda6731$ line ratio.
They found lower average density in the \sii\ emitting zone in NLS1 galaxies
than in BLS1 galaxies.
V\'{e}ron-Cetty et al. (2004) presented a high S/N optical
spectrum of the NLS1 galaxy I\,Zw\,1 and concluded that the bulk of the NLR is
unlike that of most Sy1 galaxies. It is of unusually low excitation and
dominated by lines of high critical density, while lines like 
\oiii\ $\lambda$5007
and \sii\ $\lambda\lambda$6716,6731 are weak. They infer a density of the
low-ionization part of the NLR of \n\,$=10^{6-7}$\,\cm.
\citet{lao97b} detected on the same object
very weak [\ion{C}{3}] $\lambda1907$ and [\ion{Si}{3}] $\lambda1883$ emission
in a high S/N UV spectrum, suggesting a NLR component with
\n\,$ \sim 5 \times 10^5$\,\cm. 

Rodriguez-Pascual et al. (1997) analyzed the UV properties of a sample of NLS1 galaxies.
Based on their photoionization modeling of the emission lines, they
tentatively favored a {\em low-density} BLR in NLS1 galaxies.
\citet{ferland89} presented observations of \caii\ emission lines from AGN
with strong \feii\ emission. They suggested 
that the BLR density in strong \feii\ emitters is not higher 
than in other sources, based on the ratio of the forbidden lines
relative to the \caii\ triplet.
On the other hand, other studies (e.g., Kuraszkiewicz et al. 2000; 
Wills et al. 2000; Marziani et al. 2001; Bachev et al. 2004) 
favored high-density BLRs, based on the 
\ion{Si}{3}]~$\lambda1892$/\ion{C}{3}]~$\lambda1909$ ratio
which is sensitive to density; such results are consistent with
predictions by Gaskell (1985) that UV spectra of NLS1 galaxies
would show a larger \ion{Si}{3}]~$\lambda1892$/\ion{C}{3}]~$\lambda1909$
ratio. 
More recently, in studies of the optical-UV emission-line spectra of AGN,
Marziani et al. (2001) and Bachev et al. (2004) reported indications
of a systematic increase in density toward AGN with smaller FWHMs of
the broad component of \hb\, based on the line ratio
\ion{Si}{3}]~$\lambda1892$/\ion{C}{3}]~$\lambda1909$ (they could not exclude
the alternative interpretation of varying metal abundances, though).
Comastri et al. (1998) reported detection of a deviation of the BLR Balmer
line ratio \ha/\hb\ from the recombination value, such that the ratio
was below the recombination value. They interpreted this as an indication
of high density of the BLR of the NLS1 galaxies.

Our finding is that NLS1 galaxies have lower average NLR density. Therefore, 
if NLR and BLR are closely coupled, then our results favor models
which also predict lower average BLR density in NLS1 galaxies (i.e., a stronger
scatter including objects with lower density).

\subsection{Supersolar metallicities and temperature effects}
There are several indications that NLS1 galaxies have supersolar metallicities (e.g.,
Mathur 2000; Komossa \& Mathur 2001; Nagao et al. 2002; 
Shemmer and Netzer 2002; Romano et al. 2004; Fields et al. 2005). 
We already noted in Sect.\,6 that those objects with reliable 
\oiii\ $\lambda$4363 measurements do not indicate enhanced temperatures 
in the low-density objects. However, the number of good spectra is still 
relatively small. Nagao et al. (2001) also examined the temperature-sensitive 
\oiii\ ratio for a small sample of NLS1 galaxies and concluded that they do not 
significantly deviate from BLS1 galaxies.
On the theoretical side, whether an increase in metals, and specific 
elements in particular, leads to increased heating or cooling of the gas, 
and thus an increase or decrease of its temperature, needs to be assessed
by detailed photoionization calculations. 
In general, increasing the oxygen abundance first leads to a {\em decrease} 
in temperature since oxygen is an important coolant. Such an effect would 
shift the data points in the \n\,--\,FWHM(\hbb) diagram to even lower densities 
(because of the dependence of the \sii\ ratio on temperature). 
In summary, we do not expect metallicity effects to play a dominant role 
in explaining the \n\,--\,FWHM(\hbb) diagram.

\subsection{Starburst contribution}
If we had a strong starburst contribution in a fraction of our sources, 
then this would lead to lower measured density because H\,II
regions have lower average density (e.g. Osterbrock 1989; Ho et al. 1997).
If NLS1 galaxies are "young objects" (e.g., Mathur 2000; 
Mathur, Kuraszkiewicz \& Czerny 2001; Grupe \& Mathur 2004; 
Mathur \& Grupe 2005), still in the process of growing their black holes, 
then they may possibly also show enhanced starburst activity. 
However, as already being checked in Sect.6, NLS1 galaxies, including 
{\em low-density} objects, do not show signs of stronger starburst 
contribution to their optical emission lines than BLS1 galaxies. 
We thus conclude 
that starburst effects are negligible% 
\footnote{
In a study of a sample of 74 post-starburst type\,I AGN, which underwent a
strong recent star formation epoch but stopped forming stars, \citet{zhou05}
found that more than half of them are NLS1 galaxies. We checked whether their 
sample, which also includes BLS1 galaxies, shows any special preference 
for certain 
density values, particularly whether the post-starburst NLS1 galaxies would 
preferentially have low density. 
We find that their sources 
populate similar areas in the \n\,--\,FWHM(\hbb) diagram. 
We have few post-starbursts among the low-density NLS1 galaxies in our sample, 
based on the lack of evidence 
of strong post-burst features in their host galaxies
(equivalent width of H$\delta$ absorption line EW(H$\delta$) $>$ 5 \AA\ 
for post-starbursts.}.

\subsection{NLR extent}
BLS1 galaxies would lack an observable {\em low-density} NLR component, if the 
{\em low-density} part of the NLR was selectively obscured. However,
spatially-resolved spectroscopy of the NLRs of nearby Seyfert galaxies 
shows that density declines as a function of cloud distance from the center. 
This would imply that it is the outer parts being obscured,
leading to a peculiar and unlikely geometry. 

Alternatively, lower average NLR density of the small--FWHM(\hbb) objects 
may imply that their NLRs are, on average, more extended, since density 
declines outward.  Indeed, there were early 
suggestions that the emission-line regions of NLS1 galaxies are at larger nuclear 
separations than those of BLS1 galaxies (e.g., Giannuzzo et al. 1999).
Such a NLR model is consistent with a larger BLR distance in NLS1 galaxies (e.g.,  
Wandel \& Boller 1998, Puchnarewicz et al. 2001). 

However, we note, that other parameters also play a role
in determining NLR extent. 
In particular, there appears to be a scaling between luminosity of \oiii\ and 
NLR extent (Bennert et al. 2002, Schmitt et al. 2003), such that more luminous 
objects have more extended emission-line regions. 
We checked whether our sample shows a correlation between density and
\oiii\ luminosity. No obvious trend is found.

\subsection{Fraction of matter-bounded clouds}
A substantial part of the \sii\ emission is produced in the
partially ionized zone of the NLR clouds.
{\em If} the fraction of matter-bounded clouds varies 
as a function of the distance of the NLR clouds from the nucleus,
and {\em if} the fraction of matter-bounded clouds in BLS1 galaxies
is {\em higher} at {\em larger} nuclear distances than it is in
NLS1 galaxies, then (given observations of density stratification
within the NLR from high to low density) some of the
low-density clouds in BLS1 galaxies partly escape
our measurement, with the consequence that the average
measured density in BLS1 galaxies is higher. Or, to put it the other way,
a larger scatter in the number of matter-bounded low-density clouds
in NLS1 galaxies would lead to a larger scatter in their average measured
densities.

There are no direct measurements of the fraction of matter-bounded
clouds in NLS1 galaxies and BLS1 galaxies.  However, 
several observations do indicate a wide of range of
column densities of the emission line
clouds in NLS1 galaxies: Ferland \& Persson (1989) need very high column
densities to reproduce the strength of the Calcium emission in 
objects with strong \feii\ emission, while Rodriguez-Pascual et al. (1997) 
infer a higher fraction of
matter-bounded BLR clouds in NLS1 galaxies, based on UV observations.
Regarding the NLR of NLS1 galaxies, Contini et al. (2003) require
matter-bounded clouds in order to
reproduce the strength of high-ionization iron lines.   
Photoionization models of the NLR
of NLS1 galaxies presented by Rodriguez-Ardila et al. (2000b)  
invoke a mixture of matter-bounded and ionization-bounded clouds,
the inner, high-density clouds mostly matter-bounded
and producing high-ionization lines; the outer,
low-density clouds mostly ionization-bounded and producing
mostly low-ionization lines. However, 
Rodriguez-Ardila et al. (2005) discuss a different type
of NLR models, involving shocks and photoionization,
and in that model it is shocks which produce the 
low-ionization lines.    

Future spatially resolved long-slit spectroscopy of the NLRs of nearby
NLS1 galaxies will allow a direct comparison of the density profiles of
BLS1 galaxies (e.g., Bennert et al. 2006b and references therein) 
with those of NLS1 galaxies. 
Making use of other line ratios in addition to \sii\ 
may also allow us to determine the fraction of matter-bounded clouds  
as a function of radius, and enable us to measure the NLR extent of NLS1 galaxies.

\subsection{ISM of the host galaxy}
The NLR clouds are most likely directly related to the ISM of the host
galaxy. The NLR properties then might also reflect different gas enrichment
mechanisms of the ISM. For instance, the gas could be due to local stellar
processes or transported inward from much larger scales
(or else could arise from the inner parts in case of outflowing winds,
driven by the central engine; see Sect.\,7.7).
The absence of low-density clouds in BLS1 galaxies may thus reflect
the properties of the ISM in the host galaxy.

The host galaxies of NLS1 galaxies are often spirals, but not much is 
known about their systematic properties.
There are indications that the host galaxies of NLS1 galaxies and 
BLS1 galaxies differ, particularly in the sense 
that large-scale bars are more common in NLS1 galaxies \citep{cre03,ohta07}, 
and that 
NLS1 galaxies have more grand-design dust spirals 
and a higher fraction of nuclear star-forming rings \citep{deo06}. 
Moreover, \citet{kron01} examined the host galaxies 
and found NLS1 galaxies reside in galaxies with smaller 
diameters than BLS1 galaxies. 
Previous studies on mass--luminosity/sigma relation led to 
conflicting possibilities, i.e., NLS1 galaxies are hosted by 
less luminous galaxies
(e.g., Wang \& Lu 2001; Botte et al. 2004, 2005) 
or on the contrary, more luminous galaxies for a given black hole mass
(e.g., Mathur, Kuraszkiewicz \& Czerny 2001;
Grupe \& Mathur 2004; Bian \& Zhao 2004; 
Ryan et al. 2007).  

Many of the morphological differences of the host galaxies of 
NLS1 galaxies and BLS1 galaxies 
are due to the presence or absence of a large-scale stellar bar \citep{deo06}.
Dynamical models show that a bar potential can efficiently drive gas from the 
outer regions (several kpc) to within $\sim$1\,kpc from the nucleus, at which 
point the bar-driven gas flow slows or even stalls at the inner Lindblad 
resonance and the infalling gas will form a disk \citep{shl90}. 
Gas inflow along a galactic stellar bar \citep{simkin80,shl00}, has been 
proposed to be one possible trigger of AGN activity
(e.g. Shlosman, Begeman \& Frank 1990).
However, how these processes affect the ISM density distribution
on the scales of the NLR is presently unclear.
Further studies of ISM properties of NLS1 and BLS1 galaxies on sub-kpc
scales are needed to further address this issue. 

\subsection{Outflows}
 
Lawrence et al. (1997) suggested that the density of an outflowing wind
might be an important ingredient in understanding emission-line parameters 
of AGN and correlations among them. They loosely suggested strong \feii\ 
emitters have the denser winds, in the context of a model where
parts of the BLR are mechanically heated and produce the \feii\ emission. 
We do find that BLS1 galaxies and NLS1 galaxies differ indeed 
in the density of their NLRs, 
but that it is actually the BLS1 galaxies which do harbor the 
higher density clouds.
A correlation analysis further shows that \n\ is 
anti-correlated with R4570 ($r_{\rm s}=-0.47$, $P_{\rm null}<10^{-4}$; Fig.\,8).
The {\em low-density} objects show larger-than-average R4570 compared to
the {\em high-density} NLS1 galaxies. 
The average R4570 for the {\em low-density} objects and the 
{\em high-density} NLS1 galaxies is 0.95 and 0.61, respectively.

\citet{schiano86} predicted that the average NLR density should be 
higher in very luminous objects than in lower luminosity objects.
The prediction was based on their '{\em quasar wind }' model, 
i.e., the NLR is the result of the interaction of 
AGN ionizing photons and a thermal wind on
dense, massive interstellar clouds. 
However, our result conflicts with this prediction.
The {\em low-density} regime in the \n\--FWHM \hbb\ 
diagram (Fig.\,5) is not dominated by low-luminosity systems, 
but a mixture of high- and low-luminosity NLS1 galaxies. The quasars and
low-luminosity Sy1 galaxies 
have almost the same average NLR density
of $\approx$ 290\,\cm.

There are indications that many (but not all; e.g., Xu et al. 2003; 
Williams, Mathur \& Pogge 2004) 
NLS1 galaxies accrete close to or even above the Eddington limit
(e.g., Boroson \& Green 1992; Wang et al. 1996;
Boller et al. 1996; Laor et al. 1997; 
Sulentic, Marziani \& Dultzin-Hacyan 2000; Boroson 2003; Grupe 2004; 
Grupe \& Mathur 2004; Collin \& Kawaguchi 2004).
This likely comes with the presence of strong outflows. 
If these still propagate up into the NLR, then we may expect that the
NLR gas in such objects is actually more tenuous.
Radiation-pressure driven wind models predict a decrease of accretion rate with
increasing width of the broad component of \hb\ (e.g., Nicastro 2000; 
Witt et al. 1997).
Among the NLS1 population itself, 
it should then be the objects with accretion rates closer to Eddington 
that drive the stronger winds and thus have more tenuous, {\em low-density} 
NLR components. 
In order to test this, in Fig.\,9, 
we plot Eddington ratio \LLeddbol\ as a function of density \n. We
estimate the bolometric luminosities using 
$L_{\rm bol} \approx 9\lambda L_{\lambda}(5100 \AA)$ \citep{kaspi00}, 
while the Eddington luminosities are calculated using the black hole masses
determined using the BLR radius and the velocity of the BLR gas
(e.g. Peterson 1997). We find that \LLeddbol\ is higher in 
NLS1 galaxies than in BLS1 galaxies.
An anti-correlation of decreasing electron density with
increasing Eddington ratio can be seen across our entire
sample of NLS1 and BLS1 galaxies 
($r_{\rm s}=-0.42$, $P_{\rm null}=10^{-4}$).
However, among the NLS1 population itself,
the {\em low-density} objects do not show higher-than-average 
\LLeddbol\ compared to the {\em high-density} NLS1 galaxies, which may suggest
that higher \LLeddbol\ is a necessary 
but not a sufficient 
condition to lower density. 
Independent calculation of \LLedd,
e.g., estimation from X-ray observations, 
is of great importance to check the trend.
49 out of 93 objects have X-ray counterparts in the RASS catalogs
(e.g. Voges et al. 1999). 
We had a first look at the X-ray data. 40 of them have enough photon 
counts for an estimation of the X-ray slope and thus X-ray luminosity in 
the ROSAT band. The anti-correlation between \n\ and 
\xLLedd\ is even stronger ($r_{\rm s}=-0.47$, $P_{\rm null}=3 \times 10^{-3}$),
particularly in the sense that {\em low-density} objects do have 
higher-than-average \xLLedd\ compared to the {\em high-density} NLS1 galaxies, 
with average \xLLedd\ = 3.2 and 1.2, respectively. 
Further careful study of the X-ray spectra and follow-up X-ray 
observations of the {\em low-density} objects will be 
crucial to understand the cause of the correlation. 

If outflow was a key mechanism to explain the lower average NLR density 
in NLS1 galaxies, 
then we would expect that the density (\sii\ ratio) scales with 
the \oiii\ outflow velocity (blueshift). 
We checked for both, NLS1 galaxies vs. BLS1 galaxies, and within 
the NLS1 sample 
({\em high-density} objects vs. {\em low-density} objects) whether 
the density correlates with the \oiii\ (peak) blueshift,
and only found a weak correlation 
($r_{\rm s}=-0.29$, $P_{\rm null}=7 \times 10^{-3}$)%
\footnote{This does not yet exclude that the whole NLR is in outflow. 
The \oiii\ peak blueshift was calculated relative to the
the low-ionization lines; i.e., \sii\ and \nii. Ideally,
one should measure the shift between host galaxy absorption lines and NLR
emission lines, but most of our spectra are AGN dominated with few absorption
lines detected.}. 
However, it is interesting to note that the peak blueshift of \oiii\ does
strongly correlate with \LLeddbol\
($r_{\rm s}=0.51$, $P_{\rm null}<10^{-4}$).
This correlation then indicates that outflows are more common
in objects with high accretion rates.
We also checked whether there is a correlation between density and 
the blueshift of the {\em blue wing} of \oiii%
\footnote{57 out of 93 objects clearly show blue wings.
The blueshift of the peak of the blue wing was measured 
against the peak position of the core of the [O III] line.}, since a 
preferred interpretation of blue wings is the existence of outflows
(or inflows) combined with viewing angle effects (e.g., Boroson 2005).
A correlation is seen with 
$r_{\rm s}=-0.43$ ($P_{\rm null}=1\times10^{-3}$).
This correlation shows that outflows are stronger in 
the {\em low-density} objects (Fig.\,10), 
even if the bulk of the NLR does not participate in the outflow.

In summary, we find several indications which point toward a link between
NLR density and outflows, and we tentatively favor the role of outflows 
in explaining the difference in the NLR density between
NLS1 galaxies and BLS1 galaxies. 

\section{Summary and conclusions}
We have studied one of the largest homogeneously analyzed 
sample of NLS1 galaxies 
in order to examine whether or not there is a difference in electron density
\n\ between NLS1 galaxies and BLS1 galaxies.
We employ a powerful diagnostic,  
the density-sensitive line ratio \sii\ $\lambda6716/\lambda6731$, to 
measure the NLR density. We show that the galaxies do not homogeneously 
populate the \n\,--FWHM(\hbb) diagram.
Our key finding is the detection of a
'{\em zone of avoidance}' in the \n\,--\,FWHM(\hbb) plane:
BLS1 galaxies (FWHM(\hbb) $>$ 2000 \kms) avoid low average densities,
and all show \n\,$>140$\,\cm. 
On the other hand, NLS1 galaxies show a larger scatter in densities in the range
\n\,$=2 \sim 770$\,\cm, including a significant number of objects with
low densities.
The results obtained for \sii\ are consistent with the \oi\,--\,FWHM(\hbb)
diagram, which shows higher average \oi/H$\alpha$ intensity for BLS1 galaxies.

We investigated a number of different explanations for the 
'{\em zone of avoidance}' in density. 
We find that supersolar metallicities
and temperature effects, a strong starburst contribution in NLS1 galaxies,
and the effect of NLR extent are unlikely explanations.
Consequences of the fraction of matter bounded clouds,
and different properties of the ISM in the host galaxies,
can only be further judged with  
future observations.
We find several lines of evidence that outflows
play a significant role in driving the difference in the NLR
between NLS1 galaxies and BLS1 galaxies, and favor these as explanation
for the zone of avoidance in the density-FWHM(\hbb) diagram. 

\acknowledgments
DX and SK thank the Chinese National Science
Foundation (NSF) for support under grant NSFC-10503005.
DX acknowledges the Max-Planck-Institut f\"ur extraterrestrische Physik and
the Max-Planck-Gesellschaft for financial support.
We are grateful to Hongling Lu for running her independent component 
analysis software.
We also thank Chen Cao, Biwei Jiang and Xiaobo Dong for helpful discussion 
on various softwares used in the spectral analysis, and Martin Gaskell,
Caina Hao and Jing Wang for useful conversations. 
This research made use of the SDSS archives and the Catalogue of Quasars 
and Active Nuclei. 
Funding for the
creation and the distribution of the SDSS Archive has been provided
by the Alfred P. Sloan Foundation, the Participating Institutions,
the National Aeronautics and Space Administration, the National
Science Foundation, the U.S. Department of Energy, the Japanese
Monbukagakusho, and the Max Planck Society. The SDSS is managed by
the Astrophysical Research Consortium (ARC) for the Participating
Institutions. The Participating Institutions are The University of
Chicago, Fermilab, the Institute for Advanced Study, the Japan
Participation Group, The Johns Hopkins University, Los Alamos
National Laboratory, the Max-Planck-Institute for Astronomy (MPIA),
the Max-Planck-Institute for Astrophysics (MPA), New Mexico State
University, Princeton University, the United States Naval
Observatory, and the University of Washington.

\begin{figure}
\plotone{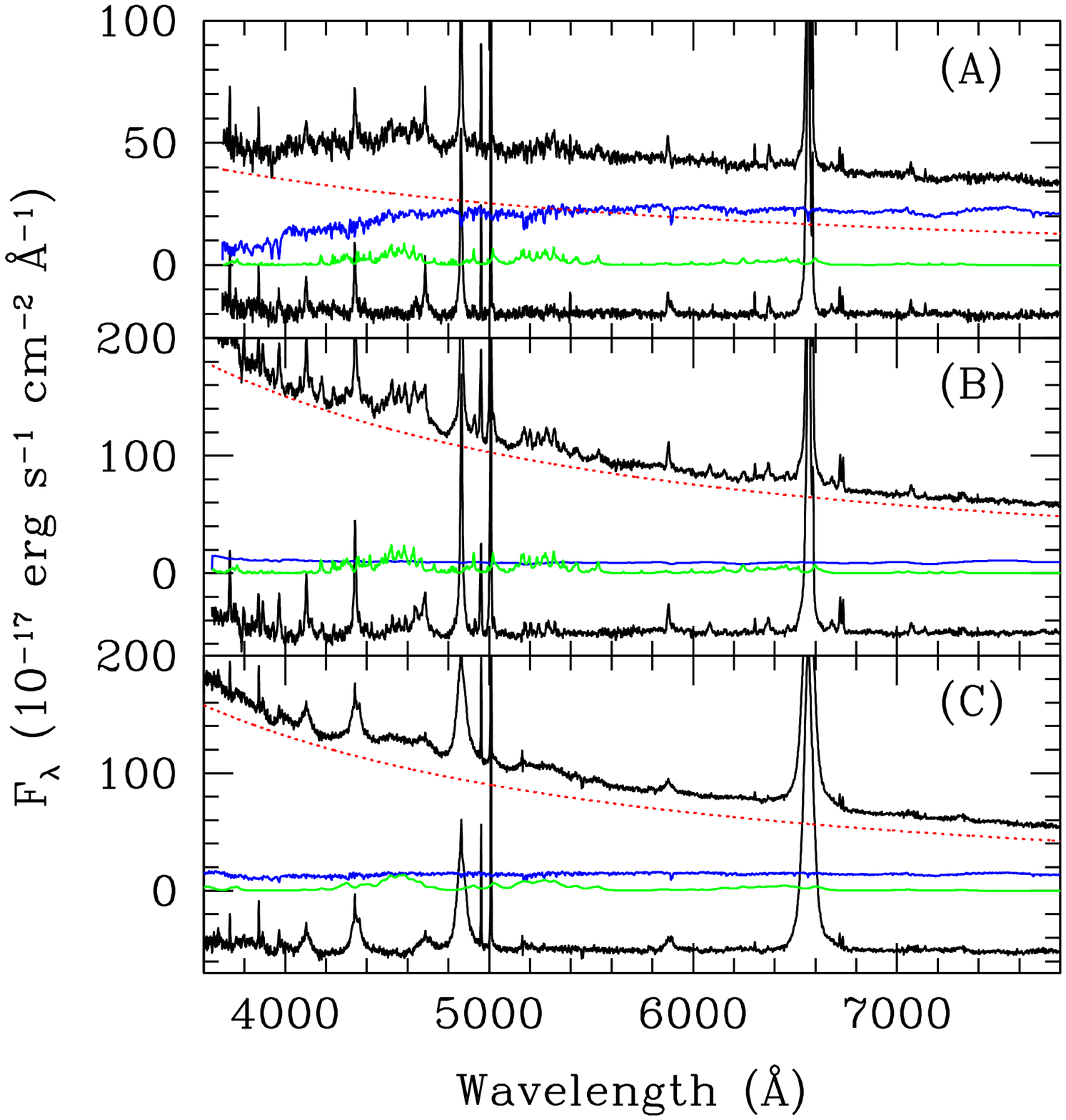}
\caption{
Examples of the decomposition of the spectrum into starlight and 
nuclear continuum.
F$_\lambda$ in units of 10$^{-17}$~erg~s$^{-1}$~cm$^{-2}$~\AA$^{-1}$
is plotted against wavelength in \AA.
In each panel, the original spectrum, the power-law continuum 
of the nucleus, the host galaxy spectrum, the \feii\ template and the 
residual spectrum are shown from top to bottom. For clarity, the residual
spectrum is offset by an additive constant. 
The three examples are drawn from the regimes A, B and C  
of Fig.\,5, defined by the width of the broad component of \hb\ and 
electron density.
\label{fig1}}
\end{figure}

\begin{figure}
\plotone{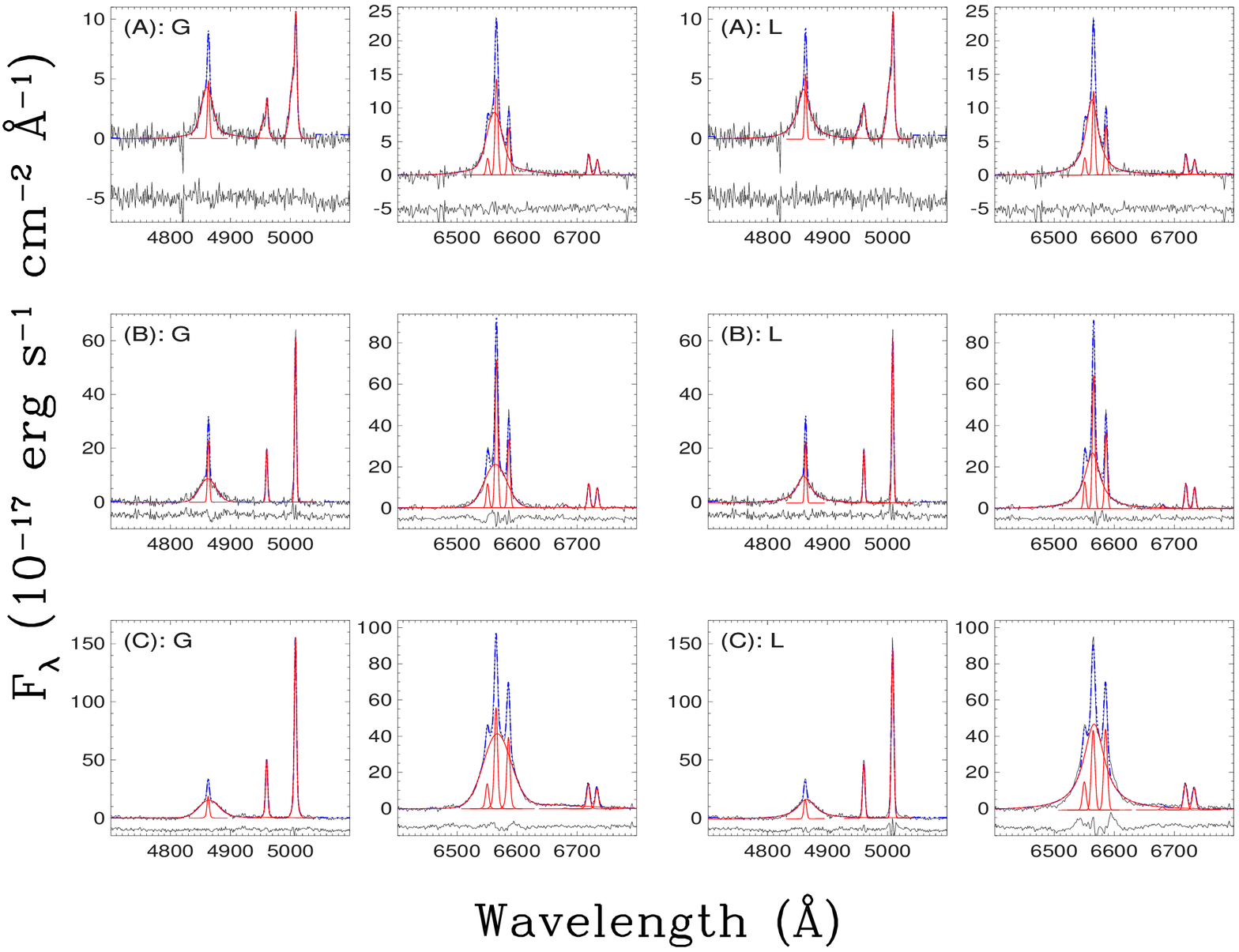}
\caption{Examples of the decompsoition of the H$\beta+$\oiii\ and 
H$\alpha+$\nii\ emission-line profiles. 
F$_\lambda$ in units of 10$^{-17}$~erg~s$^{-1}$~cm$^{-2}$~\AA$^{-1}$
is plotted against wavelength in \AA. 
The narrow emission-lines are fit by Gaussian profiles, while the 
broad Balmer components are fit by either Gaussian (G) or 
Lorentzian (L) profiles.
For Balmer lines the narrow and broad components are shown, while for 
the forbidden lines the resulting narrow line profiles are plotted. 
To illustrate the difference between the data and the fit, the resulting
residuals are shown at the bottom of each panel. 
For clarity, the residual spectrum is offset by an additive constant.
The three examples are drawn from regimes A, B and C of Fig.\,5, 
defined by the widths of \hbb\ and electron density.
\label{fig2}}
\end{figure}

\begin{figure}
\plotone{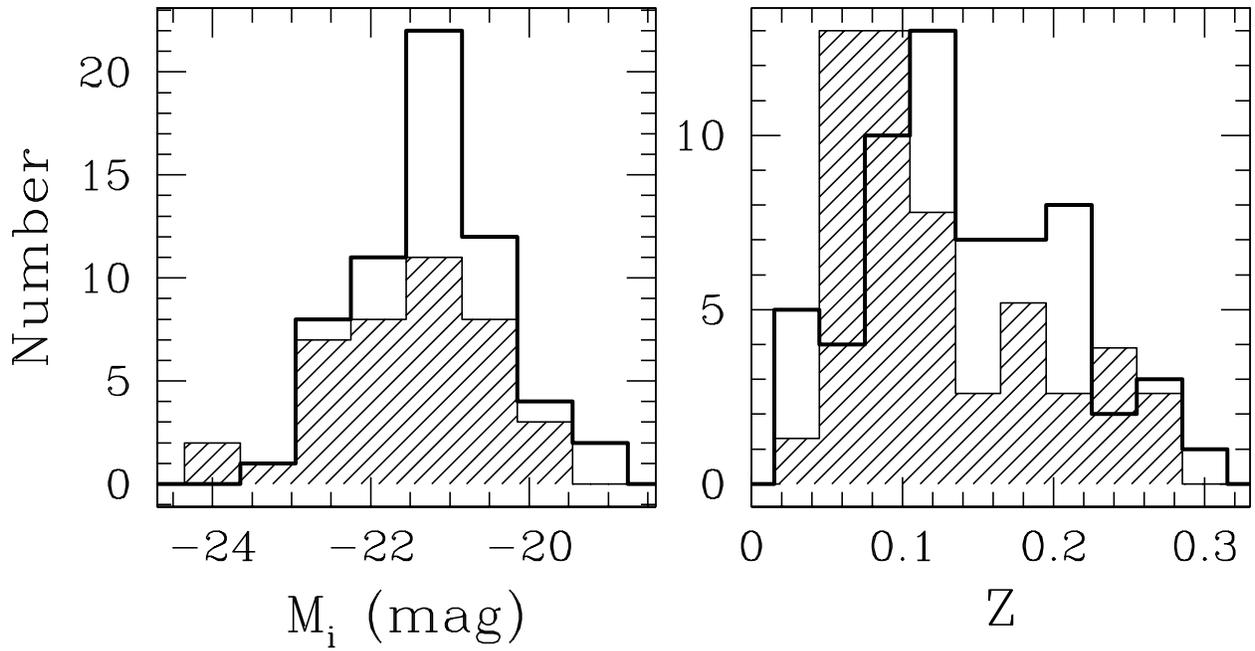}
\caption{Distributions of $i$-band absolute magnitude (left) and redshift
(right). The open histograms plot the NLS1 galaxies, and the shaded histograms
plot the BLS1 galaxies of our samples. 
\label{fig3}}
\end{figure}

\begin{figure}
\plotone{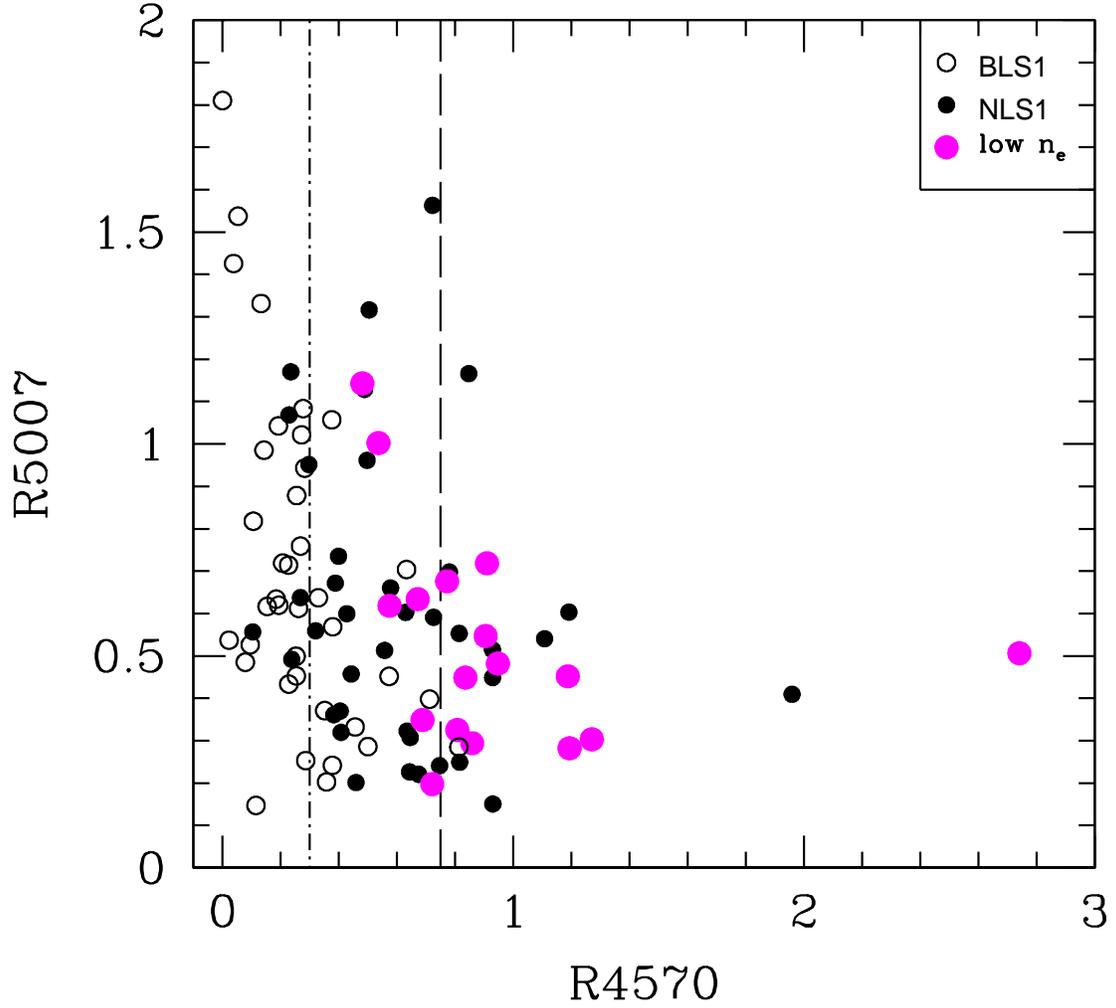}
\caption{
Ratio of \oiii\ $\lambda5007$ to the total H$\beta$ flux vs. R4570, 
the ratio of the flux of the \feii\ complex between $\lambda$4434 and 
$\lambda$4684 to that of H$\beta$ for NLS1 galaxies (filled circles) 
and BLS1 galaxies (open circles).
The dashed line and dot-dashed line 
mark the mean R4570 for NLS1 galaxies and BLS1 galaxies, 
respectively. The large filled circles (pink) represent the {\em low-density} 
objects from regime\,A of Fig.\,5. 
\label{fig4}}
\end{figure}

\begin{figure}
\plotone{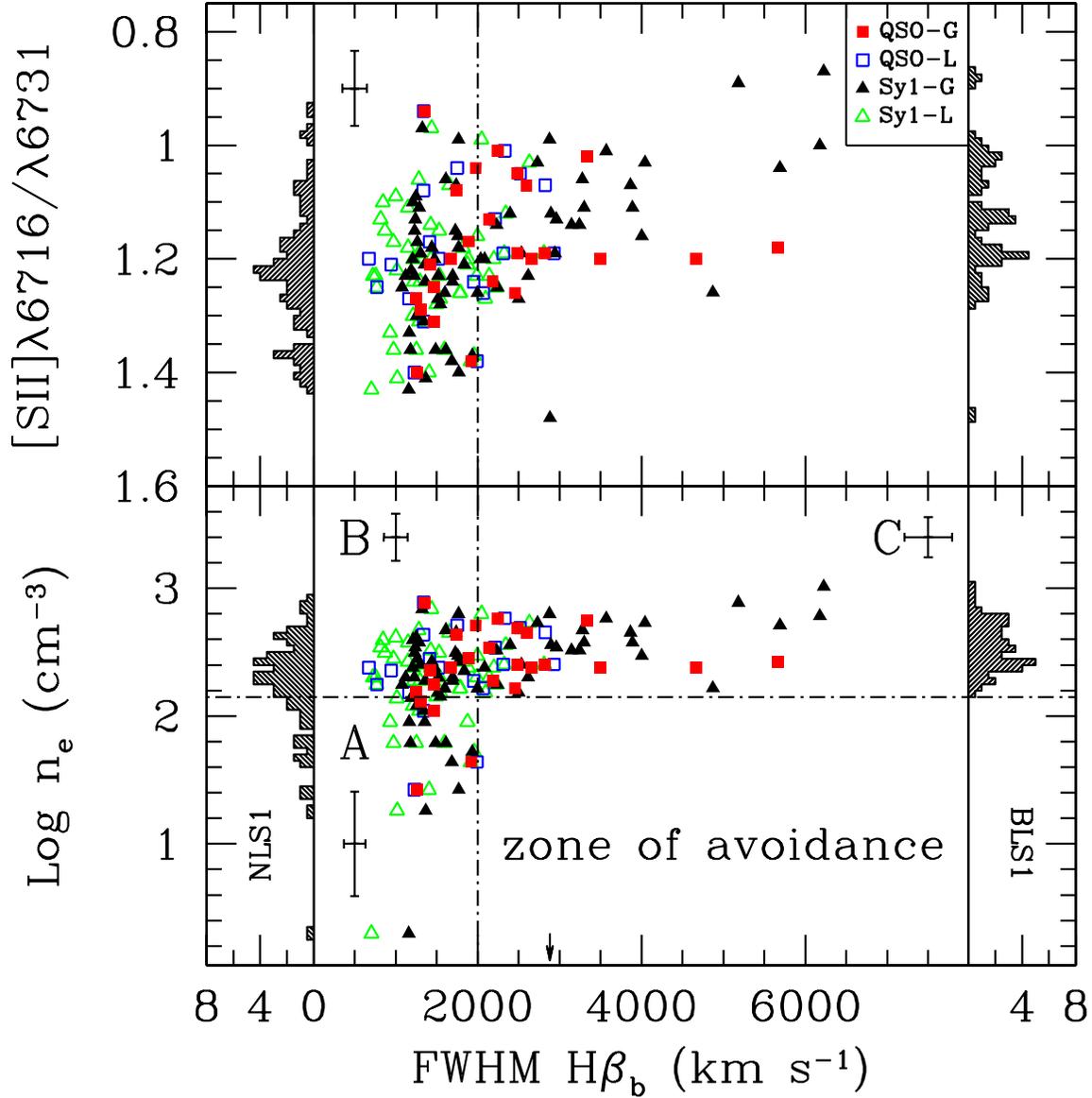}
\vspace{-1.2cm}
\caption{$\bold{Top}$: \sii\ $\lambda6716/\lambda6731$ intensity ratio
vs. FWHM of the broad component of \hb\ for our sample. 
Filled and open symbols represent the broad \hb\ components modeled by 
Gaussian (G) and Lorentzian (L) profiles, respectively. 
Squares correspond to QSOs; triangles 
to  Seyfert~1s. The median error bar is given at the upper left corner. 
The vertical dot-dashed line marks the boundary between NLS1 galaxies
and BLS1 galaxies in terms of FWHM(\hbb). 
Histograms of the \sii\ $\lambda6716/\lambda6731$ ratio 
of NLS1 galaxies and BLS1 galaxies are plotted 
in the left and right panels, respectively. 
$\bold{Bottom}$: Electron density obtained from the 
\sii\ $\lambda6716/\lambda6731$ ratio in dependence of FWHM of the broad 
component of \hb. 
Symbols as in the top panel.
The arrow points to the location of one outlier which is off the plot.
The dot-dashed lines distinguish areas populated by: 
(A) NLS1 galaxies with small width of \hbb\ and low density; 
(B) NLS1 galaxies small width of \hbb\ and high density; and 
(C) BLS1 galaxies with large width of \hbb\ and high density. 
Median error bars of each regime are given.
Distributions of the electron density of NLS1 galaxies and BLS1 galaxies 
are plotted 
in the left and right panels, respectively.
\label{fig5}}
\end{figure}

\begin{figure}
\plotone{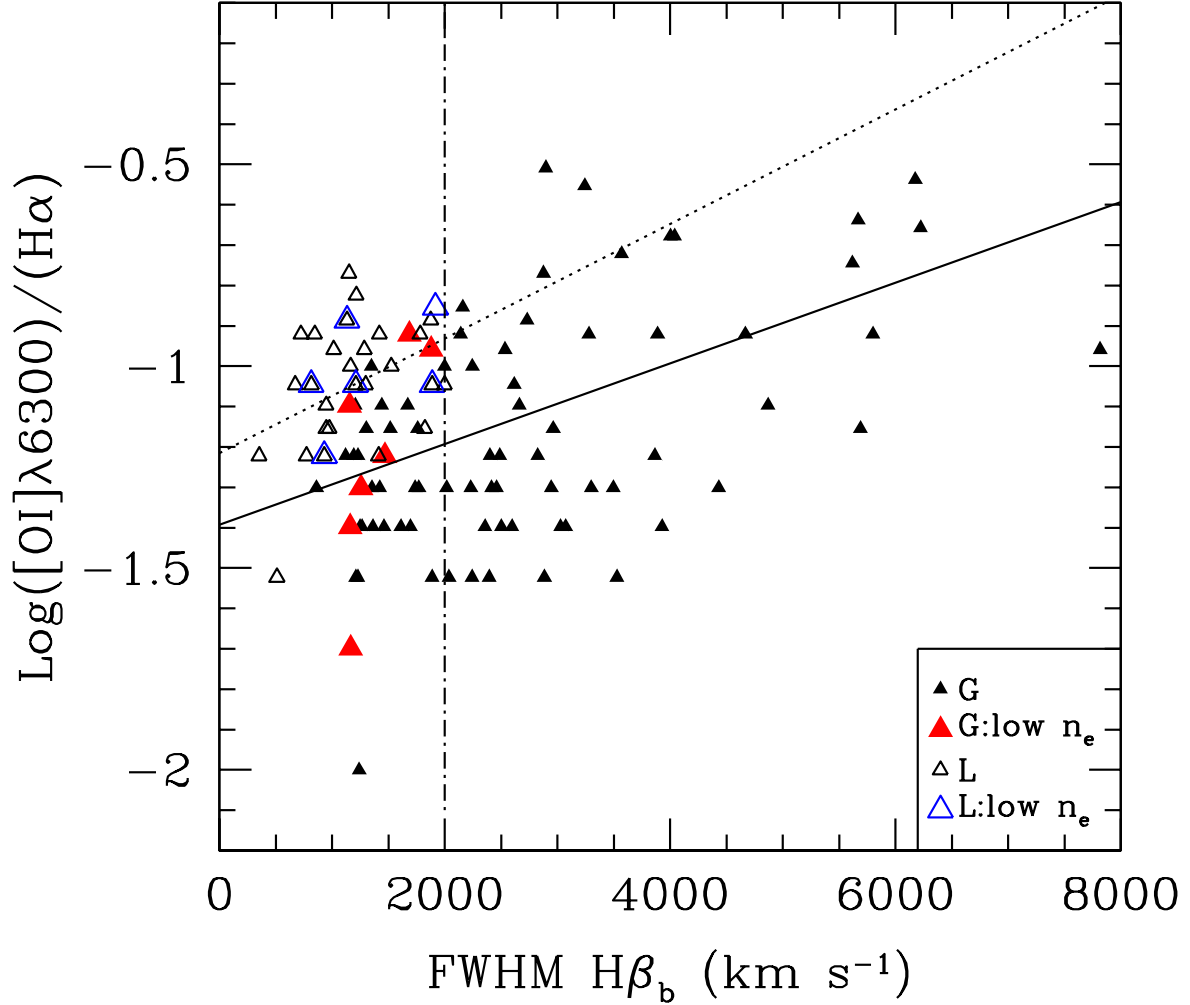}
\caption{\oi$\lambda6300$/H$\alpha_n$ ratio plotted against 
FWHM of the broad component of \hb. 
Filled triangles are data 
derived by fitting the broad Balmer components with Gaussian profiles, 
while open triangles are data derived by modeling the broad Balmer 
components with Lorentzian profiles. Large symbols represent the 
{\em low-density} objects from regime\,A of Fig.\,5. 
The \oi$\lambda6300$/H$\alpha_n$ ratio 
is correlated with FWHM(\hbb) ($r_{\rm s}=0.41$, $P_{\rm null}<10^{-3}$). 
The solid line shows the ordinary 
least-square regression fit to filled triangles, 
the dotted line the fit to open triangles. 
The dot-dashed line marks the boundary between NLS1 galaxies and BLS1 galaxies.
\label{fig6}}
\end{figure}

\begin{figure}
\plotone{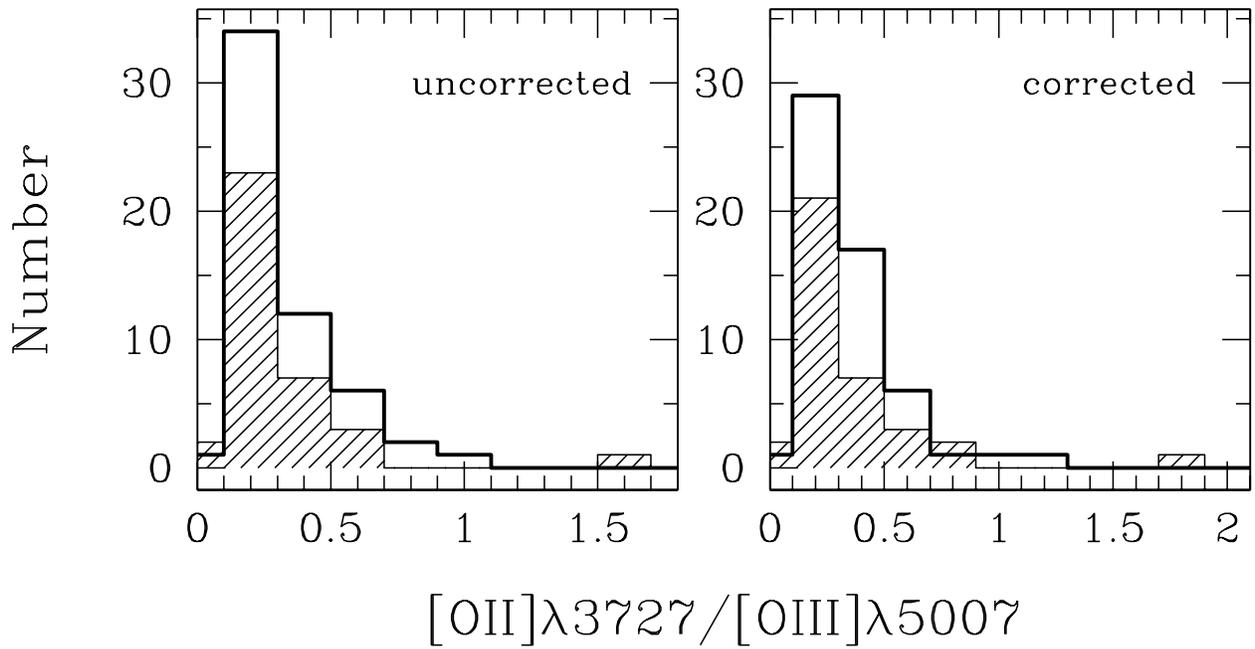}
\caption{Histograms showing the \oii\ $\lambda3727$/\oiii\ $\lambda5007$ 
distribution prior to reddening correction (left) and after reddening
correction (right). The open histograms refer to NLS1 galaxies, the shaded 
histograms to BLS1 galaxies.
\label{fig7}}
\end{figure}

\begin{figure}
\plotone{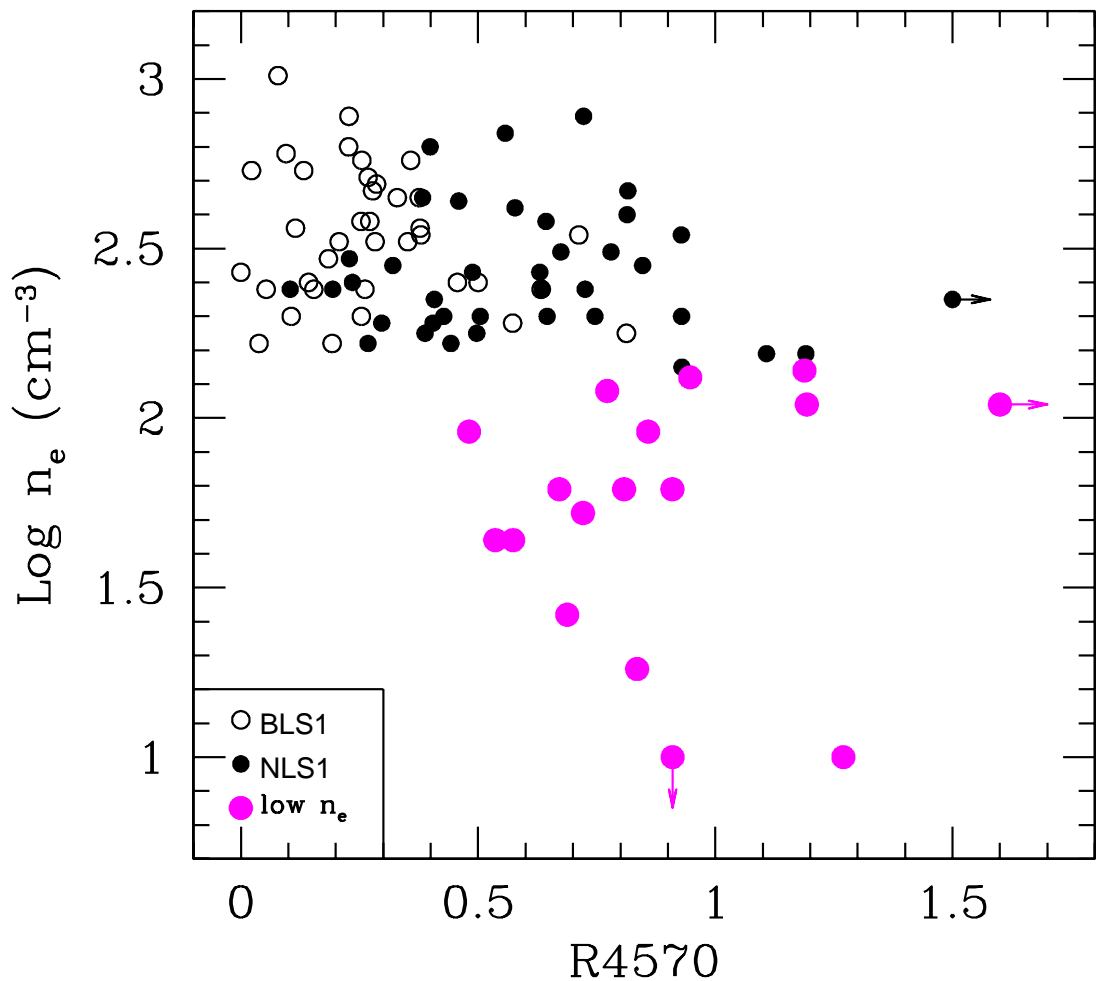}
\caption{
Electron density plotted against R4570 for NLS1 galaxies (filled circles) 
and BLS1 galaxies
(open circles). The large filled circles represent the {\em low-density}
objects from regime\,A of Fig.\,5. 
The sources which are off the plot, are indicated by arrows. 
The density is anti-correlated with R4570
($r_{\rm s}=-0.47$, $P_{\rm null}<10^{-4}$).
\label{fig8}}
\end{figure}

\begin{figure}
\plotone{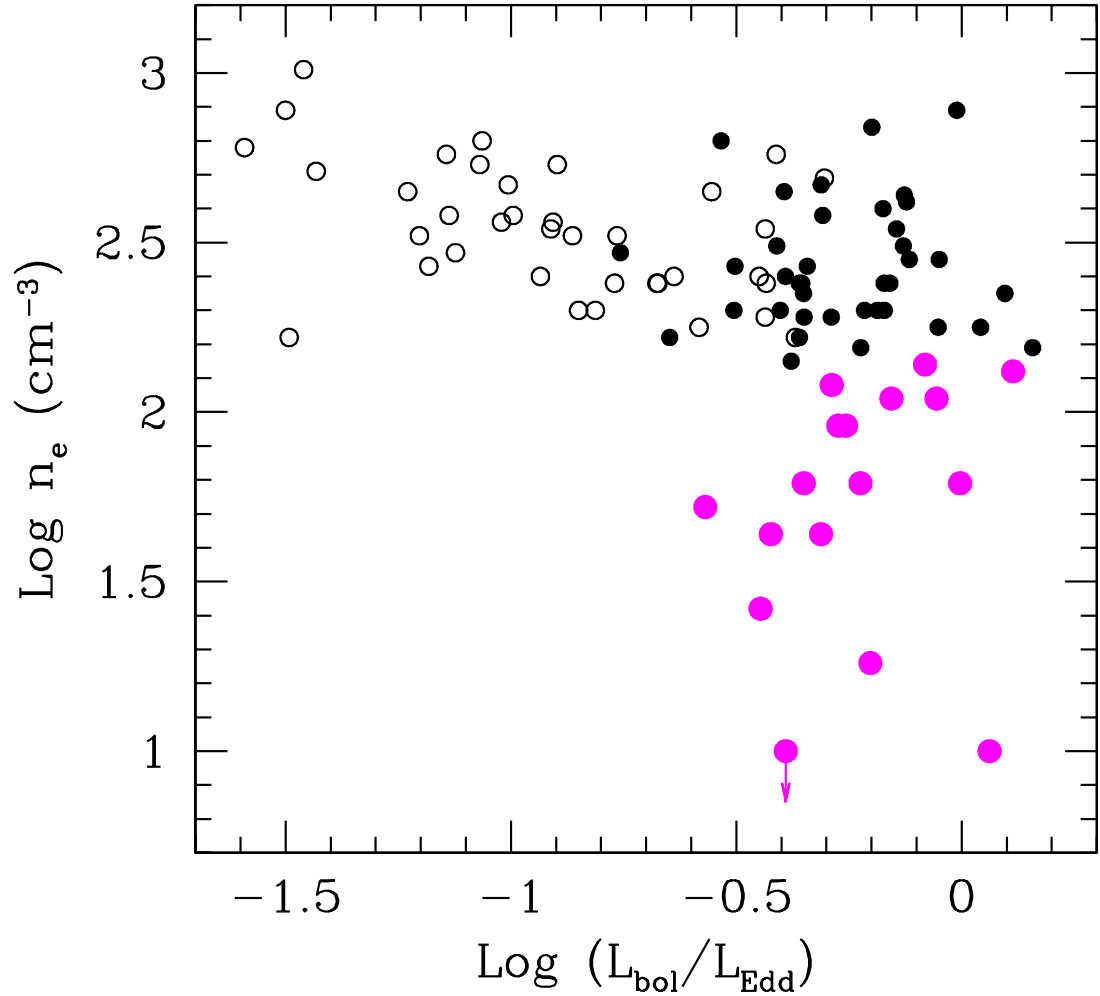}
\caption{Electron density plotted against the Eddington ratio, 
${\rm L_{bol}/L_{Edd}}$. Filled symbols correspond to NLS1 galaxies; 
open symbols to BLS1 galaxies.
Symbols are the same as in Fig.\,8.
A trend of decreasing electron density with 
increasing Eddington ratio can be 
seen ($r_{\rm s}=-0.42$, $P_{\rm null}=10^{-4}$). 
\label{fig9}}
\end{figure}

\begin{figure}
\plotone{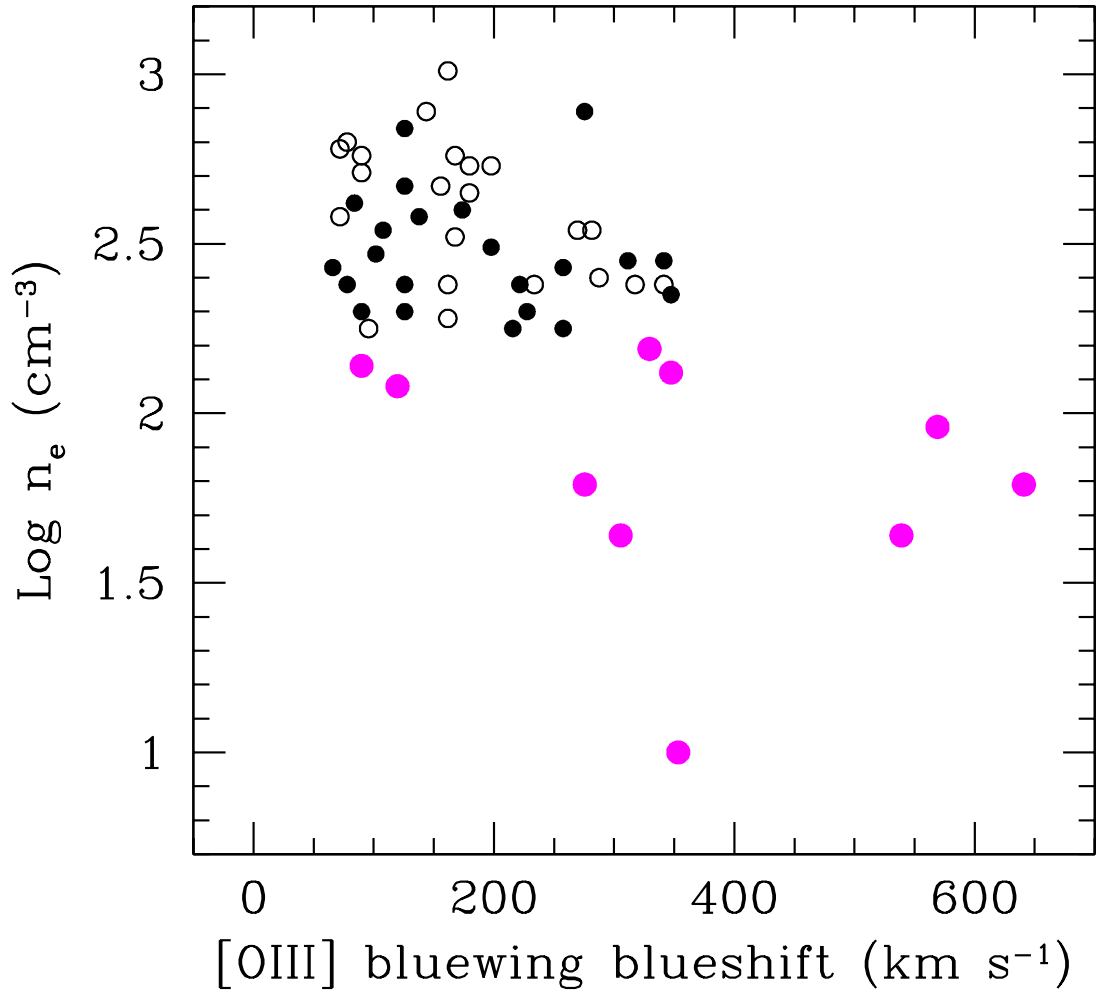}
\caption{Electron density plotted against 
the blueshift of the blue wing of \oiii. 
Symbols are the same as in Fig.\,8.
\label{fig10}}
\end{figure}

\end{document}